\documentclass[twocolumn,showpacs,amsmath,amssymb,pre,nobalancelastpage]{revtex4}

\usepackage{amsthm}
\usepackage{newlfont}
\usepackage{graphicx}
\usepackage{subfigure}
\usepackage[section]{placeins}
\usepackage{psfrag}
\usepackage{caption2}
\usepackage{flafter}
\usepackage{epsfig}
\usepackage{bm}%
\def\fnum@figure{\figurename\thefigure}
\renewcommand{\figurename}{Fig.}

\begin{document}


\title{ N-order bright and dark rogue waves in a resonant erbium-doped fibre system }
\author{Jingsong He$^{a}$}
\author{Shuwei Xu$^{b}$}
\author{K. Porseizan$^{c}$}

\address{$^{a}$Department of Mathematics, Ningbo University, Ningbo, Zhejiang 315211, P.R. China. \\$^{b}$ School of Mathematical Sciences, USTC, Hefei, Anhui 230026, P.R. China. \\$^{c}$ Department of Physics, Pondicherry University, Puducherry 605014, India.}

\begin{abstract}
The rogue waves in a resonant erbium-doped fibre system governed by
a coupled system of the  nonlinear Schr\"odinger equation and the
Maxwell-Bloch equation (NLS-MB equations)  are given explicitly by a
Taylor series expansion about the breather solutions of the
normalized slowly varying amplitude of the complex field envelope
$E$, polarization $p$ and population inversion $\eta$. The n-order breather solutions of the three fields are
constructed using Darboux transformation (DT) by assuming periodic
seed solutions. What is more, the n-order rogue waves are given by
determinant forms with $n+3$ free parameters. Furthermore, the
possible connection between our rouge waves and the generation of
supercontinuum generation is discussed.

\end{abstract}
\maketitle


\section{Introduction}

In recent years, long haul optical communication through fibers has
attracted considerable interest in research activities among
scientists all over the world.
 Especially, it has been demonstrated that the soliton-type pulse propagation will play a vital role in the ultra fast communication systems.  They are considered to be the futuristic tools in achieving low-loss, cost-effective high speed communication throughout the world.  Soliton-type pulse propagation through nonlinear optical fibers is realized by means of the exact counterbalance between the major constraints of the fiber, viz., group velocity dispersion (linear effect) which  broadens the pulse and the self-phase modulation (nonlinear effect) which contracts the pulse. The propagation of optical pulses through a nonlinear fiber in the picosecond regime is described by the well-known nonlinear Schr\"odinger (NLS) equation, which was first proposed by Hasegawa and Tappert  in 1973\cite{AB1}.

To make the soliton based communication systems highly competitive,
reliable and economical when compared to the conventional systems,
attenuation in a fiber must be compensated. A different type of optical
soliton is associated with the self-induced transparency (SIT)
effect in resonant absorbers. The soliton pulse propagation in an
erbium doped fiber amplifier utilizes the SIT phenomena, first
discovered by McCall and Hahn\cite{AB2}.  In 1967, McCall and
Hahn proposed a new type of optical soliton in a two level
resonant system. Above a well-defined threshold intensity, short
resonant pulses of a given duration will propagate through a
normally absorbing medium with anomalously low attenuation. This
happens when the pulse width is short, compared to the relaxation
times in the medium and the pulse centre frequency is in resonance
with a two-level absorbing transition. After a few classical
absorption lengths, the pulse achieves a steady state in which its
width, energy and shape remain constant. The pulse velocity has
greatly reduced from the normal velocity of light in such media.
With these properties, the pulse propagation of this type is named
"self-induced transparency" (SIT) soliton and frequently described
by the  Maxwell-Bloch (MB) equations. They are
\begin{align}
 E_z=& p, \nonumber\\
 p_t=& i\ \omega_0 p-fq\eta,\nonumber\\
 \eta_t=& f(q p^* +q^* p),\label{MB}
 \end{align}
Here, $E$ and $p$ are complex variables, $\eta$ is a real variable
and $\omega_0$ is a real constant and $f$ is the character
describing the interaction between the resonant atoms and the
optical field. The symbol * denotes the complex conjugate. These equations can be extended to the case of fiber amplifiers.
When Er is doped with the core of the optical fibres,      then the
nonlinear wave propagation can have both the effects due to silica
and Er impurities. Er impurities give SIT effect to the optical
pulse, whereas the silica material gives the NLS soliton effect.
So if we consider these effects for a large width pulse, then the system
dynamics will be governed by the  coupled system of the NLS equation
and the MB equation (NLS-MB system). Considering the Erbium doped in nonlinear silica waveguides, for the first time, the combined NLS-MB  system was proposed by Maimistov and Manykin \cite{aim1,Maimistov}in 1983. They have also constructed the Lax pair and used the inverse scattering transform technique for the generation of soliton
solution. The NLS-MB equations read as\cite{ aim1, Maimistov, aim2}

 \begin{align}
 E_t=& i[\frac{1}{2} E_{xx}+ |E|^2 E]+ 2p, \nonumber\\
 p_x=& 2i\ \omega_0 p+ 2E\eta,\nonumber\\
 \eta_x=& -(E p^* +E^* p),\label{NLS-MB}
 \end{align}
The above equation have also been reduced through the Painleve
analysis \cite{Porsezian1}. Further, Kakei and Satsuma \cite{Kakei}
also reported the Lax pair and the
 multi-soliton solution of the NLS-MB equations. The integrability aspects of NLS-MB system with variable dispersion, the study of propagation of
 optical solitons in coupled NLS-MB, and random nonuniform-doped media have been reported earlier wherein the spectral parameter was kept constant. The coexistence of NLS soliton and SIT soliton has already been confirmed experimentally\cite{mn1,mn2}. The propagation and switching of SIT in nonlinear directional couplers with two-level
 atom nonlinearity has been recently investigated numerically by retaining the transverse dependence of the optical field and atomic variable. Recent experiments by Nakazawa \emph{et al.},  have confirmed guided wave SIT soliton formation and propagation by employing a few meters of erbium doped fibre  \cite{Nakazawa1,Nakazawa2,mn1,mn2}.

Recently, considering all higher order effects in the propagation of
femtosecond pulses,  the coupled Hirota and  Maxwell-Bloch (CH-MB)
equations have been proposed and analyzed for soliton solutions
\cite{Porsezian2}.  Some generalization of NLS-MB equations, for
instance, the CH-MB equations and the NLS-MB equations with variable
dispersion and nonlinear effects are discussed
\cite{deformNLSMB,defNLSMB,deformHNLSMB}. The single soliton and the
single breather solutions\cite{he1} of the NLS-MB equations are
given by the Darboux transformation (DT)\cite{GN,matveev}.  Soliton solution for the generalized coupled variable coefficient NLS-MB system was also investigated by the DT \cite{guo111} and the Hirota method \cite{Ming}.

In recent years, in addition to solitons in different optical
systems, the study of rogue waves have also attracted considerable
interest because of their potential applications in different
branches of physics including
oceanography\cite{Kharif1,Kharif2,Didenkulova,Akhmediev2}, which
occurs due to either modulation instability
\cite{Peregrine,Dysthe1,
Zakharov2,Akhmediev3,Shrira,yan,Dai},  or
random initial condition \cite{Didenkulova,Ying}.
The first order rogue wave
 is most likely to appear as a single peak hump with two caves in a plane with
 a nonzero boundary. One of the possible generating mechanisms for  rogue waves
 is the creation of breathers which can be realized by modulation instability.  Then, larger rogue waves can
  build up when two or more breathers collide themselves\cite{Dubard1,Dubard2,
  Gaillard,guo,Yang,Akhmediev4,JSH1}. Recently, more general higher-order rogue waves
   were obtained  such as
    showing that these general N-th order rogue waves contain $(N-1)$ free irreducible
     complex
parameters\cite{Yang}. Rogue waves can also be observed in space
plasmas\cite{ruderman,derman,xuhe,xuhe2,Moslem,guoliu} and optics
when propagating high power optical radiation through photonic
crystal fibers \cite{Optical1,Optical2,Optical3}. Considering all
higher order effects in the propagation of femtosecond pulses, rogue
waves can also be observed in a system modeled by the Hirota
equation \cite{Akhmediev5,Guangye,taohe1}. Furthermore, rogue waves have not only been observed in
continuous media but have also been reported in discrete systems,
such as the systems of the well-known Ablowitz-Ladik (A-L) equation
\cite{Akhmediev6}.

Though the rogue waves have been reported in different branches of
physics where the system dynamics is governed by single equation, to
the best our knowledge, they have  been observed and reported very
little in the coupled systems. For example, Rogue waves of the
coupled NLS were constructed in the literatures
\cite{GuoBL2,Baronio,WeiGuo}. Very recently,  new kinds of matter
rogue waves \cite{Zhenyun} have been reported  in $F=1$ spinor
Bose-Einstein condensate system controlled  by three component NLS
equation. In experiment, the rogue waves in a multistable system
\cite{Rider} is revealed by experiments with an erbium-doped fiber
laser driven by harmonic pump modulation.  So, it is our prime interest to
analyze the possibility of rogue waves in coupled systems, such as
NLS-MB system.\\

It is well known that the dark soliton \cite{Zakharov} of the
defocusing nonlinear Schr\"odinger (NLS) equation is essentially
different from that of bright soliton. For the past two decades or
so, intensive research have been carried by several groups about
theoretical and experimental aspects of dark bright solitons, it is
quite natural to ask a question: is there any possibility of
observing dark rouge wave in soliton equations? In general, the
first order dark rogue wave has one down dominant peak and two small
lumps. Because of the singularity \cite{GN,matveev} of the solution
for the de-focusing NLS equation  generated by using DT, we cannot
get dark rogue wave of the de-focusing by this way. Fortunately, we
have obtained  dark  and bright rogue waves\cite{NewTypes} of the
NLS-MB equations from a Taylor series expansion of the first order
breather solutions, which are generated from a periodic seed by the
DT. But we did not provide a detailed analysis on their dynamical
evolution and higher order rogue waves. The aim of this paper is
twofold. Firstly,  the determinant representation of the n-fold DT
of the NLS-MB equations are similar to the case of NLS
equation DT\cite{he2}. Secondly, the rogue waves of the three
optical fields are constructed by determinant forms. It should be
noted that the rogue waves of the fields $p$ and $\eta$ are dark.
Furthermore, the connection between our rogue waves and the
generation of supercontinuum generation will be discussed.\\

The organization of this paper is as follows. In section 2, the
determinant representation of the n-fold DT  and formulae of
$E^{[n]}$, $p^{[n]}$ and $\eta^{[n]}$ are expressed by
eigenfunctions of spectral problem. In section 3, a Taylor series
expansion about the breather solutions are generated by n-fold DT
from a periodic seed solution with a constant amplitude to construct
the  bright and dark rogue waves. What is more, the n-order rogue
waves are given by determinant forms with $n+3$ free parameters.
Finally, we conclude the results in section 4.
\section{Darboux Transformation}

The linear spectral problem of the NLS-MB equations can be expressed
as \cite{Maimistov}
\begin{align}
\Psi_x=&U \Psi,\label{NLS-MBU}\\
\Psi_t=&V \Psi\label{NLS-MBV},
\end{align}
where
\begin{widetext}
\begin{align*}
&\Psi= \left(
\begin{array}{l}
\Psi_1\\
\Psi_2
\end{array}
\right),   \\
&U=\left[
\begin{matrix}
\lambda & E\\
-E^*     & -\lambda
\end{matrix}
\right]\equiv  \lambda \sigma_3 + U_0,    \\
&V=i \left( \left[\begin{matrix}
1& 0\\
0&-1
\end{matrix}
\right] \lambda^2 +\left[\begin{matrix}
0& E\\
-E^*& 0
\end{matrix}
\right] \lambda +\frac{1}{2}\left[\begin{matrix}
|E|^2& E_x\\
E^*_x&-|E|^2
\end{matrix}
\right]\right) +\dfrac{1}{\lambda-i\ \omega_0}\left(\begin{matrix}
\eta & -p\\
-p^* & -\eta
\end{matrix}
\right) \\
&\mbox{\hspace{.35cm}} \equiv i \sigma_3 \lambda^2 +i \lambda V_1 +
\frac{i}{2} V_{0}
        + \frac{1}{\lambda -i\ \omega_0}V_{-1},
\end{align*}
\end{widetext}
and $\lambda $ is the complex eigenvalue parameter.\\ It is easy to prove that the spectral problem (\ref{NLS-MBU}) and
(\ref{NLS-MBV}) are transformed to
  \begin{equation}\label{transform1}
{\Psi^{[1]}}_{x}=U^{[1]}~\Psi^{[1]},\ \ U^{[1]}=(T_{x}+T~U)T^{-1},
\end{equation}
\begin{equation}\label{transform2}
{\Psi^{[1]}}_{t}=V^{[1]}~\Psi^{[1]}, \ \ V^{[1]}=(T_{t}+T~V)T^{-1},
\end{equation}\\
under a gauge transformation \begin{equation}\label{transform3}
\Psi^{[1]}=T~\Psi.
\end{equation}
Here, $T$ is a $2 \times 2$ matrix, which is determined by the cross
differentiating  (\ref{transform1}) and (\ref{transform2}),
\begin{equation}\label{transformtt}
{U^{[1]}}_{t}-{V^{[1]}}_{x}+[{U^{[1]}},{V^{[1]}}]=T(U_{t}-V_{x}+[U,V])T^{-1}.
\end{equation}
This implies that, in order to make eq.(\ref{NLS-MBU}) and
eq.(\ref{NLS-MBV}) invariant under the transformation
(\ref{transform3}), it is crucial to search a matrix $T$ such that
$U^{[1]}$ and $V^{[1]}$ have the same forms as $U$ and $V$. At the
same time the old potential (or seed solutions) ($E$, $p$, $\eta$)
in spectral matrixes $U$ and $V$ are mapped into new potentials
(or new solutions) ($E^{[1]}$, $p^{[1]}$,$\eta^{[1]}$) in terms of new spectral matrixes $U^{[1]}$ and $V^{[1]}$.\\

2.1 One-fold Darboux transformation of NLS-MB equations

In order to be self-contained, we shall recall the one-fold
DT\cite{he1} of NLS-MB equations. Considering the application of the
representation for the n-fold DT by means of the
determinant\cite{he2} of  eigenfunctions with different eigenvalues
in the following context, we need to introduce $2n$ eigenfunctions
by $f_{k} = f_k(\lambda_k)=\left(
\begin{array}{c}
f_{k1}\\
f_{k2}\\
\end{array}
\right)$ associated with an eigenvalue $\lambda_{k}$,and
$\lambda_{k}=\lambda_{m}$ if $k
 = m,$ where $ k =1, 2, 3,..., 2n$ but $\lambda_{k}\neq\lambda$. Additionally, the
eigenfunctions for distinct eigenvalues are linearly independent,
(i.e.) $f_{k}$ and $f_{m}$ are linearly independent if $k\neq m$.\\

The elements of one-fold DT \cite{he1} are parameterized by the eigenfunction $f_{k}$ associated with $\lambda_k$ as
\begin{eqnarray}
T_1(\lambda;\lambda_1,\lambda_2)=\lambda I+S=\left(
\begin{array}{cc}
\dfrac{\widetilde{(T_{1})_{11}}}{|W_{2}|}& \dfrac{\widetilde{(T_{1})_{12}}}{|W_{2}|}\\ \\
\dfrac{\widetilde{(T_{1})_{21}}}{|W_{2}|}& \dfrac{\widetilde{(T_{1})_{22}}}{|W_{2}|}\\
\end{array} \right),  \label{DT1matrix}
\end{eqnarray}
Here $I$ is a unit matrix, and
\begin{eqnarray*}
S=\left(
\begin{array}{cc}
\dfrac{\begin{vmatrix}
-\lambda_{1}f_{11}&f_{12}\\
-\lambda_{2}f_{21}&f_{22}\nonumber\\
\end{vmatrix}}{|W_{2}|}& \dfrac{\begin{vmatrix}
f_{11}&-\lambda_{1}f_{11}\\
f_{21}&-\lambda_{2}f_{21}\nonumber\\
\end{vmatrix}}{|W_{2}|}\\ \\
\dfrac{\begin{vmatrix}
-\lambda_{1}f_{12}&f_{12}\\
-\lambda_{2}f_{22}&f_{22}\nonumber\\
\end{vmatrix}}{|W_{2}|}& \dfrac{\begin{vmatrix}
f_{11}&-\lambda_{1}f_{12}\\
f_{21}&-\lambda_{2}f_{22}\nonumber\\
\end{vmatrix}}{|W_{2}|}\\
\end{array} \right),
\end{eqnarray*}
\begin{equation}
W_{2}(f_{1},f_{2})=\left(\begin{matrix}
f_{11}&f_{12}\\
f_{21}&f_{22}\nonumber\\
\end{matrix}\right),det(T_{1})=(\lambda-\lambda_{1})(\lambda-\lambda_{2}),
\end{equation}
\begin{equation}
\widetilde{(T_{1})_{11}}=\begin{vmatrix}
1&0&\lambda\\
f_{11}&f_{12}&\lambda_{1}f_{11}\\
f_{21}&f_{22}&\lambda_{2}f_{21}\nonumber
\end{vmatrix},
\widetilde{(T_{1})_{12}}=\begin{vmatrix}
0&1&0\\
f_{11}&f_{12}&\lambda_{1}f_{11}\\
f_{21}&f_{22}&\lambda_{2}f_{21}\nonumber
\end{vmatrix},
\end{equation}
\begin{equation}
\widetilde{(T_{1})_{21}}=\begin{vmatrix}
1&0&0\\
f_{11}&f_{12}&\lambda_{1}f_{12}\\
f_{21}&f_{22}&\lambda_{2}f_{22}\nonumber
\end{vmatrix},
\widetilde{(T_{n})_{22}}=\begin{vmatrix}
0&1&\lambda\\
f_{11}&f_{12}&\lambda_{1}f_{12}\\
f_{21}&f_{22}&\lambda_{2}f_{22}\nonumber
\end{vmatrix}.
\end{equation}
With the transformed potentials,
\begin{eqnarray}\label{NLS-MBhe}
U_{0}^{[1]}=U_{0}-[\sigma_3,T_{1}],\nonumber\\
V_{-1}^{[1]}=T_{1}|_{\lambda=i\omega_{0}}V_{-1}{T_{1}}^{-1}|_{\lambda=i\omega_{0}},\nonumber\\
\end{eqnarray}
The  resulting new solutions of $E^{[1]}$ , $p^{[1]}$ and
$\eta^{[1]}$ are given by
\begin{widetext}
\begin{eqnarray}
&&E^{[1]}=E-2S_{12}\label{NLS-MBjix1},\\
&&p^{[1]}=-\dfrac{1}{det(T_{1})}(-2\eta(T_{1})_{11}(T_{1})_{12}+p^{\ast}(T_{1})_{12}(T_{1})_{12}-p(T_{1})_{11}(T_{1})_{11})|_{\lambda=i\omega_{0}}\label{NLS-MBjix2},\\
&&\eta^{[1]}=\dfrac{1}{det(T_{1})}(\eta((T_{1})_{11}(T_{1})_{22}+(T_{1})_{12}(T_{1})_{21})-p^{\ast}(T_{1})_{12}(T_{1})_{22}+p(T_{1})_{11}(T_{1})_{21})|_{\lambda=i\omega_{0}},\label{NLS-MBjix3}
\end{eqnarray}
\end{widetext}
and the new eigenfunction $f_k^{[1]}$ of $\lambda_k$ corresponding
to the new potentials is\begin{equation} f^{[1]}_k= \left(
\begin{array}{c}
\dfrac{\begin{vmatrix}
f_{k1}&f_{k2}&\lambda_k f_{k1}\\
f_{11}&f_{12}&\lambda_{1}f_{11}\\
f_{21}&f_{22}&\lambda_{2}f_{21}\nonumber
\end{vmatrix}}{|W_{2}|}\\
\dfrac{\begin{vmatrix}
f_{k1}&f_{k2}&\lambda_k f_{k2}\\
f_{11}&f_{12}&\lambda_{1}f_{12}\\
f_{21}&f_{22}&\lambda_{2}f_{22}\nonumber
\end{vmatrix}}{|W_{2}|}
\end{array}
\right).
\end{equation}
In order to satisfy the constraints of $S^{\prime}$ and
$V_{-1}^{\prime}$ in \cite{he1}, set
\begin{equation}\label{yuehuan1} \lambda_2 = -\lambda_1^{\ast},
f_{2}=\left(
\begin{matrix}
-f_{12}^{\ast}\\
f_{11}^{\ast}\\
\end{matrix}
\right).
\end{equation} \\

2.2 n-fold Darboux transformation for NLS-MB equations

In this subsection, our prime aim is to establish the determinant
representation of the n-fold DT for NLS-MB equations as we have done
for the case of NLS equation\cite{he2}. According to the form of $T_{1}$ in
eq.(\ref{DT1matrix}), the n-fold DT should be of the form
$T_{n}=T_{n}(\lambda)=\lambda^{n}I+t_{1} \lambda^{n-1}+t_{2}
\lambda^{n-2}+\dots+t_{n-1}\lambda+t_{n}$, where
 $t_{i}$ are 2 $\times$ 2 matrices, $i=1,2, \cdots$n. $T_{n}$ which
leads to the determinant representation of $T_n$  by means of its
kernel. Specifically, from algebraic equations,

\begin{eqnarray}
\begin{split}\label{ttnss}
f_{k}^{[n]}=T_{n}(\lambda;\lambda_1,\lambda_2,\cdots,\lambda_{2n-1},\lambda_{2n})|_{\lambda=\lambda_i}f_{k}=\\
\sum_{l=0}^{n}t_{l}\lambda_{k}^{l}f_{k}=0,
i=1,2,\cdots,2n-1,2n,
\end{split}
\end{eqnarray}
with
\begin{equation}
\label{ttnss1}t_0=\left(
\begin{array}{cc}
1&0\\
0&1\\
\end{array} \right),\nonumber\\
\end{equation}
coefficients $t_l$, $l=1,2,\dots,n$ are solved by Cramer's rule.
Thus we obtain the determinant representation of the $T_n$.\\
\noindent {\bf Theorem 1.} The n-fold DT of the NLS-MB equations is
$T_{n}=T_{n}(\lambda)=\lambda^{n}I+t_{1} \lambda^{n-1}+t_{2}
\lambda^{n-2}+\dots+t_{n-1}\lambda+t_{n}$,where $t_{i}$ are 2
$\times$ 2 matrices, $i=1,2,\cdots$ n. The final form of
$T_{n}(\lambda)$ has the form,
\begin{equation}
\label{fss1}T_{n}=T_{n}(\lambda;\lambda_1,\lambda_2,\cdots,\lambda_{2n-1},\lambda_{2n})=\left(
\begin{array}{cc}
\dfrac{\widetilde{(T_{n})_{11}}}{|W_{2n}|}& \dfrac{\widetilde{(T_{n})_{12}}}{|W_{2n}|}\\ \\
\dfrac{\widetilde{(T_{n})_{21}}}{|W_{2n}|}& \dfrac{\widetilde{(T_{n})_{22}}}{|W_{2n}|}\\
\end{array} \right),
\end{equation}
Here $I$ is a unit matrix, and expression for $t_1$ and components
of $T_n$ are given in Appendix I

It is easy to construct a simple form of the determinant of $T_{n}$
\begin{equation*}
det(T_{n})=(\lambda-\lambda_{1})(\lambda-\lambda_{2})\cdots(\lambda-\lambda_{2n-1})(\lambda-\lambda_{2n})\\
\end{equation*}
Next, we consider the transformed new solutions
($E^{[n]},p^{[n]},\eta^{[n]}$) of NLS-MB equations corresponding to
the n-fold DT.\\
\noindent {\bf Corollary] 1.}For the n-fold DT, the transformed
potentials are
\begin{eqnarray}\label{NLS-MBhen}
U_{0}^{[n]}=U_{0}-[\sigma_3,T_{n}],\nonumber\\
V_{-1}^{[n]}=T_{n}|_{\lambda=i\omega_{0}}V_{-1}{T_{n}}^{-1}|_{\lambda=i\omega_{0}},\nonumber\\
\end{eqnarray}
which leads to the new solutions $E^{[n]}$ ,$p^{[n]}$ and
$\eta^{[n]}$ of the form
\begin{widetext}
\begin{eqnarray}
&&E^{[n]}=E-2(t_{1})_{12},\label{NLS-MBjien1}\\
&&p^{[n]}=-\dfrac{1}{det(T_{n})}(-2\eta(T_{n})_{11}(T_{n})_{12}+p^{\ast}(T_{n})_{12}(T_{n})_{12}-p(T_{n})_{11}(T_{n})_{11})|_{\lambda=i\omega_{0}},\label{NLS-MBjien2}\\
&&\eta^{[n]}=\dfrac{1}{det(T_{n})}(\eta((T_{n})_{11}(T_{n})_{22}+(T_{n})_{12}(T_{n})_{21})-p^{\ast}(T_{n})_{12}(T_{n})_{22}+p(T_{n})_{11}(T_{n})_{21})|_{\lambda=i\omega_{0}},\label{NLS-MBjien3}
\end{eqnarray}

and the new eigenfunction $f_k^{[n]}$ of $\lambda_k$
is\begin{equation} f^{[n]}_k= \left(
\begin{array}{c}
\dfrac{\begin{vmatrix}
f_{k1}&f_{k2}&\lambda_{k}f_{k1}&\lambda_{k}f_{k2}&\lambda_{k}^{2}f_{k1}&\lambda_{k}^{2}f_{k2}&\ldots&\lambda_{k}^{n-1}f_{k1}&\lambda_{k}^{n-1}f_{k2}&\lambda_{k}^{n}f_{k1}\\
f_{11}&f_{12}&\lambda_{1}f_{11}&\lambda_{1}f_{12}&\lambda_{1}^{2}f_{11}&\lambda_{1}^{2}f_{12}&\ldots&\lambda_{1}^{n-1}f_{11}&\lambda_{1}^{n-1}f_{12}&\lambda_{1}^{n}f_{11}\\
f_{21}&f_{22}&\lambda_{2}f_{21}&\lambda_{2}f_{22}&\lambda_{2}^{2}f_{21}&\lambda_{2}^{2}f_{22}&\ldots&\lambda_{2}^{n-1}f_{21}&\lambda_{2}^{n-1}f_{22}&\lambda_{2}^{n}f_{21}\\
f_{31}&f_{32}&\lambda_{3}f_{31}&\lambda_{3}f_{32}&\lambda_{3}^{2}f_{31}&\lambda_{3}^{2}f_{32}&\ldots&\lambda_{3}^{n-1}f_{31}&\lambda_{3}^{n-1}f_{32}&\lambda_{3}^{n}f_{31}\\
\vdots&\vdots&\vdots&\vdots&\vdots&\vdots&\vdots&\vdots&\vdots&\vdots\\
f_{2n1}&f_{2n2}&\lambda_{2n}f_{2n1}&\lambda_{2n}f_{2n2}&\lambda_{2n}^{2}f_{2n1}&\lambda_{2n}^{2}f_{2n2}&\ldots&\lambda_{2n}^{n-1}f_{2n1}&\lambda_{2n}^{n-1}f_{2n2}&\lambda_{2n}^{n}f_{2n1}\nonumber\\
\end{vmatrix}}{|W_{2n}|}\\
\dfrac{\begin{vmatrix}
f_{k1}&f_{k2}&\lambda_{k}f_{k1}&\lambda_{k}f_{k2}&\lambda_{k}^{2}f_{k1}&\lambda_{k}^{2}f_{k2}&\ldots&\lambda_{k}^{n-1}f_{k1}&\lambda_{k}^{n-1}f_{k2}&\lambda_{k}^{n}f_{k2}\\
f_{11}&f_{12}&\lambda_{1}f_{11}&\lambda_{1}f_{12}&\lambda_{1}^{2}f_{11}&\lambda_{1}^{2}f_{12}&\ldots&\lambda_{1}^{n-1}f_{11}&\lambda_{1}^{n-1}f_{12}&\lambda_{1}^{n}f_{12}\\
f_{21}&f_{22}&\lambda_{2}f_{21}&\lambda_{2}f_{22}&\lambda_{2}^{2}f_{21}&\lambda_{2}^{2}f_{22}&\ldots&\lambda_{2}^{n-1}f_{21}&\lambda_{2}^{n-1}f_{22}&\lambda_{2}^{n}f_{22}\\
f_{31}&f_{32}&\lambda_{3}f_{31}&\lambda_{3}f_{32}&\lambda_{3}^{2}f_{31}&\lambda_{3}^{2}f_{32}&\ldots&\lambda_{3}^{n-1}f_{31}&\lambda_{3}^{n-1}f_{32}&\lambda_{3}^{n}f_{32}\\
\vdots&\vdots&\vdots&\vdots&\vdots&\vdots&\vdots&\vdots&\vdots&\vdots\\
f_{2n1}&f_{2n2}&\lambda_{2n}f_{2n1}&\lambda_{2n}f_{2n2}&\lambda_{2n}^{2}f_{2n1}&\lambda_{2n}^{2}f_{2n2}&\ldots&\lambda_{2n}^{n-1}f_{2n1}&\lambda_{2n}^{n-1}f_{2n2}&\lambda_{2n}^{n}f_{2n2}\nonumber\\
\end{vmatrix}}{|W_{2n}|}
\end{array}.
\right)
\end{equation}
\end{widetext}
Note that
\begin{equation} \label{nconstriants}
\lambda_{2k} = -\lambda_{2k-1}^{\ast}, f_{2k}=\left(
\begin{matrix}
-f_{2k-12}^{\ast}\\
f_{2k-11}^{\ast}\\
\end{matrix}
\right)
\end{equation}
in order to satisfy the constraints of DT.\\

\section{n-order bright and dark rogue waves generated by n-order breather solutions}

By using the results of DT discussed above, breather solutions of
$E$, $p$ and $\eta$ of NLS-MB equations are generated by assuming a
periodic seed solution. Then we can construct the explicit bright
and dark rogue waves of the NLS-MB equations through a Taylor series
expansion of  the breather solutions.

Substituting, $E= d \exp[i \rho ],  p=if E,  \eta=1$ into the spectral
problem eq.(\ref{NLS-MBU})
 and eq.(\ref{NLS-MBV}), and using the method of separation of variables and the superposition
principle, the eigenfunction $f_{2k-1}$ associated with
$\lambda_{2k-1}$ is given by
\begin{widetext}
\begin{eqnarray}\label{eigenfunfornonzeroseed}
\left(\mbox{\hspace{-0.2cm}} \begin{array}{c}
 f_{2k-11}(x,t,\lambda_{2k-1})\\
 f_{2k-12}(x,t,\lambda_{2k-1})\\
\end{array}\mbox{\hspace{-0.2cm}}\right)\mbox{\hspace{-0.2cm}}=\mbox{\hspace{-0.2cm}}\left(\mbox{\hspace{-0.2cm}}\begin{array}{c}
 C_1\varpi(x,t,\lambda_{2k-1})[1,2k-1]-C_2\varpi^{\ast}(x,t,-{\lambda_{2k-1}^{\ast})}[2,2k-1]\\
 C_1\varpi(x,t,\lambda_{2k-1})[2,2k-1]+C_2\varpi^{\ast}(x,t,-{\lambda_{2k-1}^{\ast})}[1,2k-1]\\
\end{array}\mbox{\hspace{-0.2cm}}\right).
\end{eqnarray}

Here
\begin{eqnarray*}
\left(\mbox{\hspace{-0.2cm}}\begin{array}{c}
 \varpi(x,t,\lambda_{2k-1})[1,2k-1]\\
 \varpi(x,t,\lambda_{2k-1})[2,2k-1]\\
\end{array}\mbox{\hspace{-0.2cm}}\right)\mbox{\hspace{-0.2cm}}=\mbox{\hspace{-0.2cm}}\left(\mbox{\hspace{-0.2cm}}\begin{array}{c}
d \exp[\frac{i}{2}\rho +i c(\lambda_{2k-1})] \\
\left[ i \left( c_1(\lambda_{2k-1})+ \frac{b}{2} \right)
-\lambda_{2k-1} \right]
\exp[-\frac{i}{2}\rho +i c(\lambda_{2k-1})] \\
\end{array}\mbox{\hspace{-0.2cm}}\right),\\
\end{eqnarray*}
\begin{eqnarray*}
\varpi(x,t,\lambda_{2k-1})= \left( \begin{array}{c}
 \varpi(x,t,\lambda_{2k-1})[1,2k-1]\\
 \varpi(x,t,\lambda_{2k-1})[2,2k-1]\\
\end{array} \right).~~~~~
\end{eqnarray*}
\end{widetext}
Note that $\varpi(x,t,\lambda_{2k-1})$  is the basic solution of the
spectral problem eq.(\ref{NLS-MBU}) and eq.(\ref{NLS-MBV}). Here
$a,b,d,t,z \in \Bbb R$, $C_1,C_2 \in \Bbb C$,
\begin{align*}
c(\lambda_{2k-1})=&c_1(\lambda_{2k-1})x +c_2(\lambda_{2k-1})t,\\
c_1(\lambda_{2k-1})=&\sqrt{d^2-(\frac{ib}{2}-\lambda_{2k-1})^2},\\
c_2(\lambda_{2k-1})=& \left( i \lambda_{2k-1}-\frac{b}{2}-\frac{if
}{\lambda_{2k-1} -i \omega_0}
\right)c_1(\lambda_{2k-1}),\\
\rho=&a\; t+ b\; x,\\
f=&\frac{1}{2}(a+ \frac{b^2}{2}-d^2),
\end{align*}
and the $(-b +2 \omega_0)f =2$ is used for $f$.

3.1 The n-order breather solutions of NLS-MB equations

For simplicity, under the condition $C_1=C_2=1$, let
$\lambda_{2k-1}=\alpha_{2k-1}+i\dfrac{b}{2}$, such that $\mbox{\rm
Im}(\dfrac{ib}{2}-\lambda_{2k-1})=0$ and
$c_1(\lambda_{2k-1})=\sqrt{d^2 -\mbox{\rm Re}^2(\lambda_{2k-1})}\in
\Bbb R $,and using eq. (\ref{nconstriants}), then substituting
eigenfunctions eq. (\ref{eigenfunfornonzeroseed}) into eq.
(\ref{fss1}), we obtain the determinant representation of DT in the
form which is discussed in Appendix II.

Using  eq. (\ref{xianxingbi}) into
eq.(\ref{NLS-MBjien1},\ref{NLS-MBjien2},\ref{NLS-MBjien3}) with the
choice in eq.(\ref{bi11}), we can construct $E^{[n]}$,$p^{[n]}$ and
$\eta^{[n]}$. For brevity, in the following,  we are giving only an
explicit expression of $E^{[1]}$ with specific parameters
$d=1,b=2,\omega_{0}=\dfrac{1}{2}$.
\begin{widetext}
\begin{eqnarray*}
&&E^{[1]}=E+4\alpha_{1}\dfrac{v_3}{v_4}\exp(i(-5t+2x)),\nonumber\\
&&v_3=\lefteqn{-\alpha_{1}\cos(2 w_1)+(2 \cos(w_1)+2
\sqrt{1-\alpha_{1}^2} \sin(w_1) \alpha_{1}+2 {\alpha_{1}} ^2
\cos(w_1)) \cosh(w_2){}}\nonumber\\&&{}+(2 i {\alpha_{1}}^2
\sin(w_1)-2 i \sin(w_1)-2 i \alpha_{1}\cos(w_1
)\sqrt{1-{\alpha_{1}}^2}) \sinh(w_2)-2 \alpha_{1}-\sqrt{1-{\alpha_{1}}^2} \sin(2 w_1)\nonumber\\&&{}+i\sqrt{1-{\alpha_{1}}^2} \sinh(2 w_2)-\alpha_{1}\cosh(2w_2),\nonumber\\
&&v_4=\lefteqn{2 {\alpha_{1}}^2\cos(2 w_1)-2 (4
\alpha_{1}\cos(w_1)+2 \sin(w_1) \sqrt{1-{\alpha_{1}}^2})
\cosh(w_2){}}\nonumber\\&&{}
-2(-1-{\alpha_{1}}^2-\cosh(2 w_2)-\sqrt{1-{\alpha_{1}}^2} \alpha_{1}\sin(2 w_1)),\nonumber\\
&&w_1=-2\sqrt{1-{\alpha_{1}}^2} x+\dfrac{12}{5}
\sqrt{1-{\alpha_{1}}^2} t,w_2=\dfrac{26}{5}\sqrt{1-{\alpha_{1}}^2}
t.\label{breatherjie1}
\end{eqnarray*}
\end{widetext}
 The dynamical evolution of $|E^{[1]}|^{2}$, $|p^{[1]}|^{2}$ and  $\eta^{[1]}$ for the parametric choice
$d=1,b=2, \omega_{0}=\dfrac{1}{2},\alpha_{1}=0.8 $ are plotted in
the Figures 1-3, which confirms the  direct verification of the
periodic as well as decaying properties of
 typical breather solutions. The breather of $E^{[1]}$ is almost same as that of the NLS
equation, which has one upper peak and two caves in each periodic
unit. On the other hand, we observe that there are two new kinds of
breathers for $p^{[1]}$ and $\eta^{[1]}$. It is interesting to note
that new breather $p^{[1]}$ admit one upper ring  and three down
peaks in each periodic unit. Whereas the new breather $\eta^{[1]}$
has two lumps and one down peak in each periodic unit. Moreover,
these two new breathers can be called as dark breathers because the
down amplitude is dominant in both the cases. The above discussed
new properties are clearly seen in Figures 1-3.\\

3.2 The first-order rogue waves generated by first-order breather
solutions above

Similarly, under the condition $C_1=C_2=1$, substituting
eigenfunctions eq.(\ref{eigenfunfornonzeroseed}) into
eqs.(\ref{NLS-MBjien1},\ref{NLS-MBjien2},\ref{NLS-MBjien3}) with
$\lambda_{1}=\alpha_{1}+i\dfrac{b}{2}$, by assuming $ \alpha_{1}
\rightarrow d(d>0)$,
 $E^{[1]}, p^{[1]}$and $\eta^{[1]}$ become
rational solutions $\{\tilde {E}^{[1]},\tilde {p}^{[1]},\tilde
{\eta}^{[1]} \}$ in the form of rogue waves \cite{NewTypes}.
When $ x \rightarrow {\infty}, \ t \rightarrow {\infty}$ in the
above expressions, after some manupulations, we find $|\tilde
{E}^{[1]}|^{2} \rightarrow d^2$, $|\tilde {p}^{[1]}|^{2} \rightarrow
\dfrac{d^2}{(\dfrac{b}{2}-\omega_{0})^2}$ and $\tilde {\eta}^{[1]}
\rightarrow 1$. In addition to the above conditions, from $|\tilde
{E}^{[1]}|^{2}_{x}=0$ and $|\tilde {E}^{[1]}|^{2}_{t}=0$, we also
observe that the maximum amplitude of $|\tilde {E}^{[1]}|^{2}$
occurs at $t = 0$ and $x=0$ and is equal to $9 d ^2$, and the
minimum amplitude of $|\tilde {E}^{[1]}|^{2}$ occurs at $t = 0$ and
$x=\pm\dfrac{\sqrt{3}}{2d}$ and is equal to $0$. By using similar
procedure discussed above, we can also obtain the extreme value of
$|\tilde {p}^{[1]}|^{2}$ and $\tilde {\eta}^{[1]}$.

Figure 4 is plotted for the rogue wave $|\tilde {E}^{[1]}|^{2}$ with
specific parameters $d=1,b=2,\omega_{0}=\dfrac{1}{2}$. From figure
4, we infer the following interesting results:  1) the $|\tilde
{E}^{[1]}|^{2} \rightarrow 1$ by  assuming $ x \rightarrow {\infty},
\ t \rightarrow {\infty}$ which gives the asymptotic plane; 2) The
maximum amplitude of $|\tilde {E}^{[1]}|^{2}$ occurs at $t = 0$ and
$x=0$ and is equal to 9, and the minimum amplitude of $|\tilde
{E}^{[1]}|^{2}$ occurs at $t = 0$ and $x=\pm\dfrac{\sqrt{3}}{2}$ and
is equal to $0$. As the general expression of the extreme values of
$|{p}^{[1]}|^{2}$ and $\tilde {\eta}^{[1]}$ are quite complicated in
nature, for simplicity, we  only discuss these solutions under
certain choice of parameters.

Figure 5 is plotted for the rogue wave  $|\tilde {p}^{[1]}|^{2}$
 on ($x-t$) plane with the above
parameters. Like in the earlier case, here also we observe the
following salient features: 1) the height of the asymptotical plane
is $4$  because $|\tilde {p}^{[1]}|^{2} \rightarrow 4$, when $ x
\rightarrow {\infty}, \ t \rightarrow {\infty}$; 2) The maximum
amplitude of $|\tilde {p}^{[1]}|^{2}$ occurs in the form of ring
curve on $(x-t)$ plane defined by $-\dfrac{507} {16}
t^2+\dfrac{1681}{8} t^4+\dfrac{25}{8}
x^4-\dfrac{55}{128}+\dfrac{65}{16} x^2+\dfrac{277}{4} x^2 t^2-15 x^3
t+\dfrac{65}{4}x t-123 x t^3=0$, and is equal to 5,
 and the minimum amplitude of $|\tilde {p}^{[1]}|^{2}$ occurs at four points
 \{ $(t=\dfrac{ +5 +\sqrt{5}}{52},x=\dfrac{+ 19 +9 \sqrt{5}}{52})$,
  $(t=\dfrac{+5 -\sqrt{5}}{52},x=\dfrac{+ 19 -9 \sqrt{5}}{52})$,
  $(t=\dfrac{- 5 +\sqrt{5}}{52},x=\dfrac{- 19 +9 \sqrt{5}}{52})$,
  $(t=\dfrac{- 5 -\sqrt{5}}{52},x=\dfrac{- 19 -9
  \sqrt{5}}{52})$\}  and is
   equal to $0$; 3) the extreme value of the amplitude $|\tilde {p}^{[1]}|^{2}$
  occurs at $t = 0$ and $x=0$ and is equal to $\dfrac{4}{25}$.  We also observe that the middle down
 peak in Figure 5 has two sub-peaks. Due to the direction of the observation of the
 figure, these two close sub-peaks are not clearly distinguished from the figure, we just find three
 down peaks.\\

Figure 6 is plotted for the rogue wave  $\tilde {\eta}^{[1]}$  with specific parameters as in fig.4.
From the Figure, we observe the following new results: 1) the height
of the asymptotical plane  is $1$ because  $\tilde {\eta}^{[1]}
\rightarrow 1$ by  letting $ x \rightarrow {\infty}, \ t \rightarrow
{\infty}$; 2)  the maximum amplitude of $\tilde {\eta}^{[1]}$ occurs
at two points \{$(t =+ \dfrac{5 + \sqrt{5}}{52}, x=+ \dfrac{ 19 + 9
\sqrt{5}}{52})$  \} and \{$(t =- \dfrac{5 + \sqrt{5}}{52}, x=-
\dfrac{ 19 + 9 \sqrt{5}}{52})$ \} and is equal to $\sqrt{5}$, and
the minimum amplitude of $\tilde {\eta}^{[1]}$ occurs at two points
\{$(t = \dfrac{+ 5 - \sqrt{5}}{52}, x= \dfrac{ + 19 - 9
\sqrt{5}}{52})$  \} and \{$(t = \dfrac{- 5 + \sqrt{5}}{52}, x=
\dfrac{ - 19 + 9 \sqrt{5}}{52})$ \} and is equal to $-\sqrt{5}$; 3)
the extreme value of the amplitude $\tilde {\eta}^{[1]}$ occurs at
$t = 0$ and $x=0$ and is equal to $-\dfrac{11}{5}$.  Like in
Figure.5,here also we observe that the down peak in Figure 6 has two
sub-peaks.\\

3.3 The higher order rogue waves and their determinant forms

In order to emphasize the richness of the higher order rogue waves, we can modify
 $C_1$ and $C_2$ in the equation (\ref{eigenfunfornonzeroseed}) as the following:
\begin{eqnarray}
&&C_1=K_0+\exp(ic_1(\lambda_{2k-1})\sum_{j=0}^{k-1}J_j(\lambda_{2k-1}-(d+i\dfrac{b}{2}))^{j})\nonumber\\
&&C_2=K_0+\exp(-ic_1(\lambda_{2k-1})\sum_{j=0}^{k-1}J_j(\lambda_{2k-1}-(d+i\dfrac{b}{2}))^{j})\nonumber\\
\label{chooseCC}
\end{eqnarray}
Here $K_0,J_j \in \Bbb C$. Note that
$\lambda_{2k-1}=d+i\dfrac{b}{2}$ is the zero point of
$c_1(\lambda_{2k-1})$.

Based on the section 3.2, higher order rogue waves can be
constructed by the breather solutions. In other words, let $
\lambda_{2k-1}\rightarrow d+i\dfrac{b}{2}$ in n-order breather
solutions, n-order rogue waves can be given. Generally, in comparison to
the method of limiting the breather solutions, the method of making
rational eigenfunction below may be more direct and the rogue wave
can be shown by determinant forms.

Substituting eq.(\ref{chooseCC}) into
eqs.(\ref{eigenfunfornonzeroseed}), by assuming $
\lambda_{2k-1}\rightarrow d+i\dfrac{b}{2}$,
  eigenfunction $f_{2k-1}$ associated with
$\lambda_{2k-1}$ become rational eigenfunction $f_r$ as follows.
\begin{widetext}
\begin{eqnarray*}
\left(\mbox{\hspace{-0.2cm}} \begin{array}{c}
 f_{r1}\\
 f_{r2}\\
\end{array}\mbox{\hspace{-0.2cm}}\right)\mbox{\hspace{-0.2cm}}=\mbox{\hspace{-0.2cm}}\left(\mbox{\hspace{-0.2cm}}\begin{array}{c}
 -\mbox{\hspace{-0.1cm}}((2dK_0+2d)x\mbox{\hspace{-0.1cm}}+\mbox{\hspace{-0.1cm}}2(\dfrac{2i}{(b-2\omega_0)(d+\dfrac{1}{2}ib-i\omega_0)}\mbox{\hspace{-0.1cm}}+\mbox{\hspace{-0.1cm}}id-b)(K_0+1)dt+2
J_0d+K_0+1)\sqrt{2d}\exp(K)\\
 -\mbox{\hspace{-0.1cm}}(-(2dK_0+2d)x\mbox{\hspace{-0.1cm}}-\mbox{\hspace{-0.1cm}}2(\dfrac{2i}{(b-2\omega_0)(d+\dfrac{1}{2}ib-i\omega_0)}\mbox{\hspace{-0.1cm}}+\mbox{\hspace{-0.1cm}}id-b)(K_0+1)dt-2
J_0d+K_0+1)\sqrt{2d}\exp(-K)\\
\end{array}\mbox{\hspace{-0.2cm}}\right),
\end{eqnarray*}
\begin{eqnarray}\label{reigenfunfornonzeroseed}
K=\dfrac{1}{2}i(-\dfrac{8+b^3-2b^2\omega_0-2d^2b+4
d^2\omega_0}{2(b-2\omega_0)}t+bx)
\end{eqnarray}
\end{widetext}

 Substituting
eigenfunctions eq.(\ref{reigenfunfornonzeroseed}) into
eqs.(\ref{NLS-MBjix1},\ref{NLS-MBjix2},\ref{NLS-MBjix3}), we can get
the first order rogue waves  $\{\bar {E}^{[1]},\bar {p}^{[1]},\bar
{\eta}^{[1]} \}$ in the form of determinant.  The dynamical
evolution of $|\bar {E}^{[1]}|^{2}$, $|\bar{p}^{[1]}|^{2}$ and
$\bar{\eta}^{[1]}$ for the parametric choice $d, b,
\omega_{0}, K_0, J_0$ are respectively similar to the Figures 4-6, but we can control the position of the
first-order rogue waves by choosing the parameters $K_0$ and $J_0$.

\noindent {\bf Theorem 2.} For the n-fold DT,  the n-order rogue
waves $\bar {E}^{[n]},\bar {p}^{[n]}$ and $\bar {\eta}^{[n]}$ of the
form
\begin{widetext}
\begin{eqnarray}
&&\bar {E}^{[n]}=E-2(\bar{t}_{r1})_{12},\label{rNLS-MBjien1}\\
&&\bar {p}^{[n]}=-\dfrac{1}{det(\bar{T}_{rn})}(-2\eta(\bar{T}_{rn})_{11}(\bar{T}_{rn})_{12}+p^{\ast}(\bar{T}_{rn})_{12}(\bar{T}_{rn})_{12}-p(\bar{T}_{rn})_{11}(\bar{T}_{rn})_{11})|_{\lambda=i\omega_{0}},\label{rNLS-MBjien2}\\
&&\bar
{\eta}^{[n]}\mbox{\hspace{-0.1cm}}=\mbox{\hspace{-0.2cm}}\dfrac{1}{det(\bar{T}_{rn})}(\eta((\bar{T}_{rn})_{11}(\bar{T}_{rn})_{22}\mbox{\hspace{-0.1cm}}+\mbox{\hspace{-0.1cm}}(\bar{T}_{rn})_{12}(\bar{T}_{rn})_{21})\mbox{\hspace{-0.1cm}}-\mbox{\hspace{-0.1cm}}p^{\ast}(\bar{T}_{rn})_{12}(\bar{T}_{rn})_{22}\mbox{\hspace{-0.1cm}}+\mbox{\hspace{-0.1cm}}p(\bar{T}_{rn})_{11}(\bar{T}_{rn})_{21})|_{\lambda=i\omega_{0}},\label{rNLS-MBjien3}
\end{eqnarray}
\end{widetext}
The final form of $\bar{T}_{rn}(\lambda)$ is given in Appendix III.

\noindent {\bf Case 1).}  When $n=2$, substituting eq.(\ref{rfss1})
into eq.(\ref{rNLS-MBjien1}), eq.(\ref{rNLS-MBjien2}) and
eq.(\ref{rNLS-MBjien3}) can give the second-order rogue waves with
five free parameters.  Note that under the condition $J_1>>J_0$, the
second-rogue can split into three first-order rogue wave (triplets
rogue wave) \cite{triplets} rather than two. The dynamical evolution
of $|\bar {E}^{[2]}|^{2}$, $|\bar{p}^{[2]}|^{2}$ and
$\bar{\eta}^{[2]}$ for the parametric choice $d=1,b=2,
\omega_{0}=\dfrac{1}{2},K_0=1,J_0=0,J_1=100 $ are plotted in the
Figures 7, 9 and 11 and their corresponding density plots are shown in the Figures 8, 10 and 12. There is another kind of
second-order rogue wave,for example, $|\bar {E}^{[2]}|^{2}$ is
higher than second-rogue above. The dynamical evolution of $|\bar
{E}^{[2]}|^{2}$, $|\bar{p}^{[2]}|^{2}$ and $\bar{\eta}^{[2]}$ for
the parametric choice $d=2,b=0,
\omega_{0}=\dfrac{1}{2},K_0=1,J_0=0,J_1=0 $ are plotted in the
Figures 13-15.   Note that eigenvalue
$\lambda_1=\lambda_3$ is real. The eigenvalue of rogue waves are
different from the eigenvalue of solutions given in the past.

\noindent {\bf Case 2).}  When $n=3$, substituting eq.(\ref{rfss1})
into eq.(\ref{rNLS-MBjien1}), eq.(\ref{rNLS-MBjien2}) and
eq.(\ref{rNLS-MBjien3}) can give the third-order rogue waves with
six free parameters. Note that under the condition $J_2>>J_i
 (i=0,1)$  or $J_1>>J_i  (i=0,2)$, the third-rogue can split into six
first-order rogue wave rather. Circular rogue wave \cite{Circular}
may be constructed by the condition $J_2>>J_1$ and $J_2>>J_0$. The
dynamical evolution of $|\bar {E}^{[3]}|^{2}$, $|\bar{p}^{[3]}|^{2}$
and $\bar{\eta}^{[3]}$ for the parametric choice $d=1,b=2,
\omega_{0}=\dfrac{1}{2},K_0=1,J_0=0,J_1=0, J_2=8000 $ are plotted in
the Figures 16-18.   At the same time,
triplets rogue wave  may be constructed by the condition $J_1>>J_2$
and $J_1>>J_0$. The dynamical evolution of $|\bar {E}^{[3]}|^{2}$,
$|\bar{p}^{[3]}|^{2}$ and $\bar{\eta}^{[3]}$ for the parametric
choice $d=1,b=2, \omega_{0}=\dfrac{1}{2},K_0=1,J_0=0,J_1=100, J_2=0
$ are plotted in the Figures 19-21.
Similarly, there is another kind third-order rogue wave, for
example, $|\bar {E}^{[3]}|^{2}$ is higher than third-rogue above.
The dynamical evolution of $|\bar {E}^{[3]}|^{2}$,
$|\bar{p}^{[3]}|^{2}$ and $\bar{\eta}^{[3]}$ for the parametric
choice $d=\dfrac{4}{3},b=0,
\omega_{0}=\dfrac{1}{2},K_0=1,J_0=0,J_1=0,J_2=0 $ are plotted in the
Figures 22-24.  Note that eigenvalue
$\lambda_1=\lambda_3=\lambda_5$ is real. The eigenvalue of rogue
waves are different from the eigenvalue of solutions given in the
past. According to analysis above, the n-order rogue waves may be
controlled by $n+3$ free parameters.

From the above discussions, it is interesting to point out that the
down amplitudes are dominant in the profile of rogue waves $\tilde
{p}^{[1]}$ and  $\tilde {\eta}^{[1]}$, so they are new kind of rogue
waves when compared with the typical bright rogue wave $\tilde
{E}^{[1]}$, which  are corresponding to the dark breathers in Figure
2 and Figure 3. So from our earlier understanding of breathers in
other physical systems, we call these new type of solutions as dark
rogue waves. Moreover, the dark rogue wave $\tilde {p}^{[1]}$ has
one upper ring  and three down peaks, and dark rogue wave  $\tilde
{\eta}^{[1]}$ has two lumps and one down peak.  According to
analysis, the n-order rogue waves must be generated by n-order
breather solution.

From the detailed literature on rogue waves, to the best of our
knowledge, so far only bright rogue waves have been analyzed in
detail but there is little report about the dark rogue waves in
physical system. In the case of bright optical rogue waves, many
results have actually connected the generation of supercontinuum
generation (SCG) with rogue waves\cite {scg}. In recent years, the
supercontinuum white coherent source has attracted a lot of
attention because of its potential applications in optical coherence
tomography, spectroscopy, wavelength division multiplexing, etc. As
reported in ref.\cite {kpMI}, the modulational instability (MI)
conditions for the generation of ultra-short pulses has already been
investigated in the erbium doped nonlinear fibre and occurrences of
nonconventional side bands have also been observed. This type of
nonconventional side bands will be very useful to generate large MI
bandwidth which intern generates very short pulses. In this
way,
 we believe that our rogue wave results in this paper can also be
  connected to the generation of SCG. Similarly, the occurrence of
  dark rogue
 wave can also be connected to the results of ref.\cite {kpMI},
  in the following manner: In our above work\cite {kpMI}, it has been shown
 that both bright and dark SIT solitons can be generated in the case of the anomalous and normal group velocity dispersion (GVD), in contrast
 to the well-established results in the conventional fibre, where bright and dark solitons exists in the anomalous and normal GVD regions,
  respectively. From the above results, it is clear that the formation of dark rogue waves can also be connected in a similar way. Thus,
 it is interesting to analyze the relation connecting the MI, SCG and rogue wave formation in
 optical system.

\section{Conclusion}

Thus, in this article, we have reported the  rogue waves
of the three physical fields $E, p$, and $\eta$ in a resonant
erbium-doped fibre system, which is governed by the NLS-MB
equations. These rogue waves are constructed by a Taylor series
expansion of the corresponding breather solutions of the NLS-MB
equations.  As expected, in contrast to the usual bright rogue wave
$E$, we observe dark rogue waves for $p$ and $\eta$. The main
feature of the dark rogue waves is the appearance of two (or more)
dominant down peaks in its profile. In particular, there is one
upper ring in the profile of the $p$, so it may be called as dark
ring-rogue wave.  The explicit form of $\bar {E}^{[n]},\bar
{p}^{[n]}$ and $\bar {\eta}^{[n]}$ are given by the determinant
representation of the n-fold DT. The rogue waves in previous section
can also be connected to the supercontinuum generation.

As we have already described in the introduction,  the
singularity\cite{GN} of the solutions generated by the DT  is the
main constraint to generate the dark rogue waves of the defocusing
NLS equation. This perhaps shows that the dark rogue wave of the
defocusing NLS equation can be investigated by other way such as
by means of Hirota method. From the determinant
representation of the $T_n$, it is interesting to generate the
higher order rogue waves so that the dynamical interactions of rogue
waves can be analysed. \\

In recent years, considering variable dispersion,
variable nonlinearity and variable gain/loss, the
investigation of solitons in nonautonomous nonlinear
 evolution equation equations has also attracted a lot of
  attention among researchers \cite{serkin1, serkin2,serkin3, kp1}.
   For example, Serkin and his coworkers have proposed a novel
   method to analyse the nonautonomous soliton
   equations\cite{serkin1, serkin2,serkin3} and the interaction
   of solitons in variable coefficient higher order NLS equation
   have been investigated in detail\cite{kp1}. Using the results of
    the above papers and making use of our results in this paper,
     one can also construct the multi solitons, breathers and rogue
     wave solutions of the variable coefficient NLS-MB system.
Moreover, it is also possible to obtain new type of rogue waves for
other important coupled system in optics, such as the
CH-MB equations \cite{JSH2} and variable coefficient CH-MB
equations\cite{liheuu,shanxue}.

{\bf Acknowledgments} {\noindent \small  This work is supported by
the NSF of China under Grant No.10971109 and No. 11271210 and
K.C.Wong Magna Fund in Ningbo University. Jingsong He is also
supported the Natural Science Foundation of Ningbo under Grant
No.2011A610179. KP wishes to thank the DST, DAE-BRNS, and UGC,
Government of India, for the financial support through major
projects. We thank Prof. Yishen Li(USTC,Hefei, China) for his useful
suggestions on the rogue wave. }

\section{\textbf{Appendices}}
\begin{center}
\appendix{\textbf{Appendix I}}:
In this appendix, we are giving expression for t$_1$ and
 elements of ${T_n}$
\end{center}
\begin{widetext}
\begin{eqnarray*}
t_{1}=\left(
\begin{array}{cc}
\dfrac{\widetilde{(Q_{n})_{11}}}{|W_{2n}|}& \dfrac{\widetilde{(Q_{n})_{12}}}{|W_{2n}|}\\ \\
\dfrac{\widetilde{(Q_{n})_{21}}}{|W_{2n}|}& \dfrac{\widetilde{(Q_{n})_{22}}}{|W_{2n}|}\\
\end{array} \right),
\end{eqnarray*}
\begin{equation*}\label{fsst1}
W_{2n}=\left(\begin{matrix}
f_{11}&f_{12}&\lambda_{1}f_{11}&\lambda_{1}f_{12}&{\lambda_{1}}^2 f_{11}&{\lambda_{1}}^2 f_{12}&\ldots&{\lambda_{1}}^{n-1} f_{11}&{\lambda_{1}}^{n-1} f_{12}\\
f_{21}&f_{22}&\lambda_{2}f_{21}&\lambda_{2}f_{22}&{\lambda_{2}}^2 f_{21}&{\lambda_{2}}^2 f_{22}&\ldots&{\lambda_{2}}^{n-1} f_{21}&{\lambda_{2}}^{n-1} f_{22}\\
f_{31}&f_{32}&\lambda_{3}f_{31}&\lambda_{3}f_{32}&{\lambda_{3}}^2 f_{31}&{\lambda_{3}}^2 f_{32}&\ldots&{\lambda_{3}}^{n-1} f_{31}&{\lambda_{3}}^{n-1} f_{32}\\
\vdots&\vdots&\vdots&\vdots&\vdots&\vdots&\vdots&\vdots&\vdots\\
f_{2n1}&f_{2n2}&\lambda_{2n}f_{2n1}&\lambda_{2n}f_{2n2}&{\lambda_{2n}}^2 f_{2n1}&{\lambda_{2n}}^2 f_{2n2}&\ldots&{\lambda_{2n}}^{n-1} f_{2n1}&{\lambda_{2n}}^{n-1} f_{2n2}\nonumber\\
\end{matrix}\right),
\end{equation*}
\begin{equation*}\label{fsst2}
\widetilde{(T_{n})_{11}}=\begin{vmatrix}
1&0&\lambda&0&\lambda^{2}&0&\ldots&\lambda^{n-1}&0&\lambda^{n}\\
f_{11}&f_{12}&\lambda_{1}f_{11}&\lambda_{1}f_{12}&\lambda_{1}^{2}f_{11}&\lambda_{1}^{2}f_{12}&\ldots&\lambda_{1}^{n-1}f_{11}&\lambda_{1}^{n-1}f_{12}&\lambda_{1}^{n}f_{11}\\
f_{21}&f_{22}&\lambda_{2}f_{21}&\lambda_{2}f_{22}&\lambda_{2}^{2}f_{21}&\lambda_{2}^{2}f_{22}&\ldots&\lambda_{2}^{n-1}f_{21}&\lambda_{2}^{n-1}f_{22}&\lambda_{2}^{n}f_{21}\\
f_{31}&f_{32}&\lambda_{3}f_{31}&\lambda_{3}f_{32}&\lambda_{3}^{2}f_{31}&\lambda_{3}^{2}f_{32}&\ldots&\lambda_{3}^{n-1}f_{31}&\lambda_{3}^{n-1}f_{32}&\lambda_{3}^{n}f_{31}\\
\vdots&\vdots&\vdots&\vdots&\vdots&\vdots&\vdots&\vdots&\vdots&\vdots\\
f_{2n1}&f_{2n2}&\lambda_{2n}f_{2n1}&\lambda_{2n}f_{2n2}&\lambda_{2n}^{2}f_{2n1}&\lambda_{2n}^{2}f_{2n2}&\ldots&\lambda_{2n}^{n-1}f_{2n1}&\lambda_{2n}^{n-1}f_{2n2}&\lambda_{2n}^{n}f_{2n1}\nonumber\\
\end{vmatrix},
\end{equation*}
\begin{equation*}\label{fsst3}
\widetilde{(T_{n})_{12}}=\begin{vmatrix}
0&1&0&\lambda&0&\lambda^{2}&\ldots&0&\lambda^{n-1}&0\\
f_{11}&f_{12}&\lambda_{1}f_{11}&\lambda_{1}f_{12}&\lambda_{1}^{2}f_{11}&\lambda_{1}^{2}f_{12}&\ldots&\lambda_{1}^{n-1}f_{11}&\lambda_{1}^{n-1}f_{12}&\lambda_{1}^{n}f_{11}\\
f_{21}&f_{22}&\lambda_{2}f_{21}&\lambda_{2}f_{22}&\lambda_{2}^{2}f_{21}&\lambda_{2}^{2}f_{22}&\ldots&\lambda_{2}^{n-1}f_{21}&\lambda_{2}^{n-1}f_{22}&\lambda_{2}^{n}f_{21}\\
f_{31}&f_{32}&\lambda_{3}f_{31}&\lambda_{3}f_{32}&\lambda_{3}^{2}f_{31}&\lambda_{3}^{2}f_{32}&\ldots&\lambda_{3}^{n-1}f_{31}&\lambda_{3}^{n-1}f_{32}&\lambda_{3}^{n}f_{31}\\
\vdots&\vdots&\vdots&\vdots&\vdots&\vdots&\vdots&\vdots&\vdots&\vdots\\
f_{2n1}&f_{2n2}&\lambda_{2n}f_{2n1}&\lambda_{2n}f_{2n2}&\lambda_{2n}^{2}f_{2n1}&\lambda_{2n}^{2}f_{2n2}&\ldots&\lambda_{2n}^{n-1}f_{2n1}&\lambda_{2n}^{n-1}f_{2n2}&\lambda_{2n}^{n}f_{2n1}\nonumber\\
\end{vmatrix},
\end{equation*}
\begin{equation*}\label{fsst4}
\widetilde{(T_{n})_{21}}=\begin{vmatrix}
1&0&\lambda&0&\lambda^{2}&0&\ldots&\lambda^{n-1}&0&0\\
f_{11}&f_{12}&\lambda_{1}f_{11}&\lambda_{1}f_{12}&\lambda_{1}^{2}f_{11}&\lambda_{1}^{2}f_{12}&\ldots&\lambda_{1}^{n-1}f_{11}&\lambda_{1}^{n-1}f_{12}&\lambda_{1}^{n}f_{12}\\
f_{21}&f_{22}&\lambda_{2}f_{21}&\lambda_{2}f_{22}&\lambda_{2}^{2}f_{21}&\lambda_{2}^{2}f_{22}&\ldots&\lambda_{2}^{n-1}f_{21}&\lambda_{2}^{n-1}f_{22}&\lambda_{2}^{n}f_{22}\\
f_{31}&f_{32}&\lambda_{3}f_{31}&\lambda_{3}f_{32}&\lambda_{3}^{2}f_{31}&\lambda_{3}^{2}f_{32}&\ldots&\lambda_{3}^{n-1}f_{31}&\lambda_{3}^{n-1}f_{32}&\lambda_{3}^{n}f_{32}\\
\vdots&\vdots&\vdots&\vdots&\vdots&\vdots&\vdots&\vdots&\vdots&\vdots\\
f_{2n1}&f_{2n2}&\lambda_{2n}f_{2n1}&\lambda_{2n}f_{2n2}&\lambda_{2n}^{2}f_{2n1}&\lambda_{2n}^{2}f_{2n2}&\ldots&\lambda_{2n}^{n-1}f_{2n1}&\lambda_{2n}^{n-1}f_{2n2}&\lambda_{2n}^{n}f_{2n2}\nonumber\\
\end{vmatrix},
\end{equation*}
\begin{equation*}\label{fsst5}
\widetilde{(T_{n})_{22}}=\begin{vmatrix}
0&1&0&\lambda&0&\lambda^{2}&\ldots&0&\lambda^{n-1}&\lambda^{n}\\
f_{11}&f_{12}&\lambda_{1}f_{11}&\lambda_{1}f_{12}&\lambda_{1}^{2}f_{11}&\lambda_{1}^{2}f_{12}&\ldots&\lambda_{1}^{n-1}f_{11}&\lambda_{1}^{n-1}f_{12}&\lambda_{1}^{n}f_{12}\\
f_{21}&f_{22}&\lambda_{2}f_{21}&\lambda_{2}f_{22}&\lambda_{2}^{2}f_{21}&\lambda_{2}^{2}f_{22}&\ldots&\lambda_{2}^{n-1}f_{21}&\lambda_{2}^{n-1}f_{22}&\lambda_{2}^{n}f_{22}\\
f_{31}&f_{32}&\lambda_{3}f_{31}&\lambda_{3}f_{32}&\lambda_{3}^{2}f_{31}&\lambda_{3}^{2}f_{32}&\ldots&\lambda_{3}^{n-1}f_{31}&\lambda_{3}^{n-1}f_{32}&\lambda_{3}^{n}f_{32}\\
\vdots&\vdots&\vdots&\vdots&\vdots&\vdots&\vdots&\vdots&\vdots&\vdots\\
f_{2n1}&f_{2n2}&\lambda_{2n}f_{2n1}&\lambda_{2n}f_{2n2}&\lambda_{2n}^{2}f_{2n1}&\lambda_{2n}^{2}f_{2n2}&\ldots&\lambda_{2n}^{n-1}f_{2n1}&\lambda_{2n}^{n-1}f_{2n2}&\lambda_{2n}^{n}f_{2n2}\nonumber\\
\end{vmatrix},
\end{equation*}
\begin{equation*}
\widetilde{(Q_{n})_{11}}=\begin{vmatrix}
f_{11}&f_{12}&\lambda_{1}f_{11}&\lambda_{1}f_{12}&{\lambda_{1}}^2 f_{11}&{\lambda_{1}}^2 f_{12}&\ldots&{\lambda_{1}}^{n-1} f_{12}&{\lambda_{1}}^{n} f_{11}\\
f_{21}&f_{22}&\lambda_{2}f_{21}&\lambda_{2}f_{22}&{\lambda_{2}}^2 f_{21}&{\lambda_{2}}^2 f_{22}&\ldots&{\lambda_{2}}^{n-1} f_{22}&{\lambda_{2}}^{n} f_{21}\\
f_{31}&f_{32}&\lambda_{3}f_{31}&\lambda_{3}f_{32}&{\lambda_{3}}^2 f_{31}&{\lambda_{3}}^2 f_{32}&\ldots&{\lambda_{3}}^{n-1} f_{32}&{\lambda_{3}}^{n} f_{31}\\
\vdots&\vdots&\vdots&\vdots&\vdots&\vdots&\vdots&\vdots&\vdots\\
f_{2n1}&f_{2n2}&\lambda_{2n}f_{2n1}&\lambda_{2n}f_{2n2}&{\lambda_{2n}}^2 f_{2n1}&{\lambda_{2n}}^2 f_{2n2}&\ldots&{\lambda_{2n}}^{n-1} f_{2n2}&{\lambda_{2n}}^{n} f_{2n1}\nonumber\\
\end{vmatrix},
\end{equation*}
\begin{equation*}
\widetilde{(Q_{n})_{12}}=-\begin{vmatrix}
f_{11}&f_{12}&\lambda_{1}f_{11}&\lambda_{1}f_{12}&{\lambda_{1}}^2 f_{11}&{\lambda_{1}}^2 f_{12}&\ldots&{\lambda_{1}}^{n-1} f_{11}&{\lambda_{1}}^{n} f_{11}\\
f_{21}&f_{22}&\lambda_{2}f_{21}&\lambda_{2}f_{22}&{\lambda_{2}}^2 f_{21}&{\lambda_{2}}^2 f_{22}&\ldots&{\lambda_{2}}^{n-1} f_{21}&{\lambda_{2}}^{n} f_{21}\\
f_{31}&f_{32}&\lambda_{3}f_{31}&\lambda_{3}f_{32}&{\lambda_{3}}^2 f_{31}&{\lambda_{3}}^2 f_{32}&\ldots&{\lambda_{3}}^{n-1} f_{31}&{\lambda_{3}}^{n} f_{31}\\
\vdots&\vdots&\vdots&\vdots&\vdots&\vdots&\vdots&\vdots&\vdots\\
f_{2n1}&f_{2n2}&\lambda_{2n}f_{2n1}&\lambda_{2n}f_{2n2}&{\lambda_{2n}}^2 f_{2n1}&{\lambda_{2n}}^2 f_{2n2}&\ldots&{\lambda_{2n}}^{n-1} f_{2n1}&{\lambda_{2n}}^{n} f_{2n1}\nonumber\\
\end{vmatrix},
\end{equation*}
\begin{equation*}
\widetilde{(Q_{n})_{21}}=\begin{vmatrix}
f_{11}&f_{12}&\lambda_{1}f_{11}&\lambda_{1}f_{12}&{\lambda_{1}}^2 f_{11}&{\lambda_{1}}^2 f_{12}&\ldots&{\lambda_{1}}^{n-1} f_{12}&{\lambda_{1}}^{n} f_{12}\\
f_{21}&f_{22}&\lambda_{2}f_{21}&\lambda_{2}f_{22}&{\lambda_{2}}^2 f_{21}&{\lambda_{2}}^2 f_{22}&\ldots&{\lambda_{2}}^{n-1} f_{22}&{\lambda_{2}}^{n} f_{22}\\
f_{31}&f_{32}&\lambda_{3}f_{31}&\lambda_{3}f_{32}&{\lambda_{3}}^2 f_{31}&{\lambda_{3}}^2 f_{32}&\ldots&{\lambda_{3}}^{n-1} f_{32}&{\lambda_{3}}^{n} f_{32}\\
\vdots&\vdots&\vdots&\vdots&\vdots&\vdots&\vdots&\vdots&\vdots\\
f_{2n1}&f_{2n2}&\lambda_{2n}f_{2n1}&\lambda_{2n}f_{2n2}&{\lambda_{2n}}^2 f_{2n1}&{\lambda_{2n}}^2 f_{2n2}&\ldots&{\lambda_{2n}}^{n-1} f_{2n2}&{\lambda_{2n}}^{n} f_{2n2}\nonumber\\
\end{vmatrix},
\end{equation*}
\begin{equation*}
\widetilde{(Q_{n})_{22}}=-\begin{vmatrix}
f_{11}&f_{12}&\lambda_{1}f_{11}&\lambda_{1}f_{12}&{\lambda_{1}}^2 f_{11}&{\lambda_{1}}^2 f_{12}&\ldots&{\lambda_{1}}^{n-1} f_{11}&{\lambda_{1}}^{n} f_{12}\\
f_{21}&f_{22}&\lambda_{2}f_{21}&\lambda_{2}f_{22}&{\lambda_{2}}^2 f_{21}&{\lambda_{2}}^2 f_{22}&\ldots&{\lambda_{2}}^{n-1} f_{21}&{\lambda_{2}}^{n} f_{22}\\
f_{31}&f_{32}&\lambda_{3}f_{31}&\lambda_{3}f_{32}&{\lambda_{3}}^2 f_{31}&{\lambda_{3}}^2 f_{32}&\ldots&{\lambda_{3}}^{n-1} f_{31}&{\lambda_{3}}^{n} f_{32}\\
\vdots&\vdots&\vdots&\vdots&\vdots&\vdots&\vdots&\vdots&\vdots\\
f_{2n1}&f_{2n2}&\lambda_{2n}f_{2n1}&\lambda_{2n}f_{2n2}&{\lambda_{2n}}^2 f_{2n1}&{\lambda_{2n}}^2 f_{2n2}&\ldots&{\lambda_{2n}}^{n-1} f_{2n1}&{\lambda_{2n}}^{n} f_{2n2}\nonumber\\
\end{vmatrix}.
\end{equation*}

\begin{center}
\appendix{\textbf{Appendix II: Determinant representation of Nth order DT is constructed in the form}}
\end{center}
\setcounter{equation}{0}
\renewcommand{\theequation}{A.\arabic{equation}}
\begin{equation}\label{xianxingbi}
T_{n}=T_{n}(\lambda;\lambda_1,\lambda_2,\cdots,\lambda_{2n-1},\lambda_{2n})=\left(
\begin{array}{cc}
\dfrac{\widehat{(T_{n})_{11}}}{|\bar{W}_{2n}|}& \dfrac{\widehat{(T_{n})_{12}}}{|\bar{W}_{2n}|}\\ \\
\dfrac{\widehat{(T_{n})_{21}}}{|\bar{W}_{2n}|}& \dfrac{\widehat{(T_{n})_{22}}}{|\bar{W}_{2n}|}\\
\end{array} \right),
\end{equation}
with
\begin{eqnarray*}
t_{1}=\left(
\begin{array}{cc}
\dfrac{\widehat{(Q_{n})_{11}}}{|\bar{W}_{2n}|}& -\dfrac{\widehat{(Q_{n})_{12}}}{|\bar{W}_{2n}|}\\ \\
\dfrac{\widehat{(Q_{n})_{21}}}{|\bar{W}_{2n}|}& -\dfrac{\widehat{(Q_{n})_{22}}}{|\bar{W}_{2n}|}\\
\end{array} \right),
\end{eqnarray*}
\begin{equation*}\label{fsst1}
\mbox{\hspace{-0.8cm}}\bar{W}_{2n}=\left(\begin{matrix}
1&\gamma_{1}&\lambda_{1}&\lambda_{1}\gamma_{1}&{\lambda_{1}}^2 &\ldots&{\lambda_{1}}^{n-1} &{\lambda_{1}}^{n-1} \gamma_{1}\\
-{\gamma_{1}}^{\ast}&1&{\lambda_{1}}^{\ast}{\gamma_{1}}^{\ast}&-{\lambda_{1}}^{\ast}&-{{\lambda_{1}}^{\ast}}^2{\gamma_{1}}^{\ast}&\ldots&-{(-{\lambda_{1}}^{\ast}})^{n-1}{\gamma_{1}}^{\ast}&{(-{\lambda_{1}}^{\ast})}^{n-1}\\
1&\gamma_{3}&\lambda_{3}&\lambda_{3}\gamma_{3}&{\lambda_{3}}^2 &\ldots&{\lambda_{3}}^{n-1} &{\lambda_{3}}^{n-1} \gamma_{3}\\
\vdots&\vdots&\vdots&\vdots&\vdots&\vdots&\vdots&\vdots\\
-{\gamma_{2n-1}}^{\ast}&1&{\lambda_{2n-1}}^{\ast}{\gamma_{2n-1}}^{\ast}&-{\lambda_{2n-1}}^{\ast}&-{{\lambda_{2n-1}}^{\ast}}^2{\gamma_{2n-1}}^{\ast}&\ldots&-{(-{\lambda_{2n-1}}^{\ast}})^{n-1}{\gamma_{2n-1}}^{\ast}&{(-{\lambda_{2n-1}}^{\ast})}^{n-1}\nonumber\\
\end{matrix}\right),
\end{equation*}
\begin{equation*}\label{fsst2}
\widehat{(T_{n})_{11}}=\begin{vmatrix}
1&0&\lambda&\ldots&\lambda^{n-1}&0&\lambda^{n}\\
1&\gamma_{1}&\lambda_{1}&\ldots&\lambda_{1}^{n-1}&\lambda_{1}^{n-1}\gamma_{1}&\lambda_{1}^{n}\\
-{\gamma_{1}}^{\ast}&1&{\lambda_{1}}^{\ast}{\gamma_{1}}^{\ast}&\ldots&-(-{\lambda_{1}}^{\ast})^{n-1}{\gamma_{1}}^{\ast}&(-{\lambda_{1}}^{\ast})^{n-1}&-(-{\lambda_{1}}^{\ast})^{n}{\gamma_{1}}^{\ast}\\
1&\gamma_{3}&\lambda_{3}&\ldots&\lambda_{3}^{n-1}&\lambda_{3}^{n-1}\gamma_{3}&\lambda_{3}^{n}\\
\vdots&\vdots&\vdots&\vdots&\vdots&\vdots&\vdots\\
-{\gamma_{2n-1}}^{\ast}&1&{\lambda_{2n-1}}^{\ast}{\gamma_{2n-1}}^{\ast}&\ldots&-(-{\lambda_{2n-1}}^{\ast})^{n-1}{\gamma_{2n-1}}^{\ast}&(-{\lambda_{2n-1}}^{\ast})^{n-1}&-(-{\lambda_{2n-1}}^{\ast})^{n}{\gamma_{2n-1}}^{\ast}\nonumber\\
\end{vmatrix},
\end{equation*}
\begin{equation*}\label{fsst3}
\widehat{(T_{n})_{12}}=\begin{vmatrix}
0&1&0&\ldots&0&\lambda^{n-1}&0\\
1&\gamma_{1}&\lambda_{1}&\ldots&\lambda_{1}^{n-1}&\lambda_{1}^{n-1}\gamma_{1}&\lambda_{1}^{n}\\
-{\gamma_{1}}^{\ast}&1&{\lambda_{1}}^{\ast}{\gamma_{1}}^{\ast}&\ldots&-(-{\lambda_{1}}^{\ast})^{n-1}{\gamma_{1}}^{\ast}&(-{\lambda_{1}}^{\ast})^{n-1}&-(-{\lambda_{1}}^{\ast})^{n}{\gamma_{1}}^{\ast}\\
1&\gamma_{3}&\lambda_{3}&\ldots&\lambda_{3}^{n-1}&\lambda_{3}^{n-1}\gamma_{3}&\lambda_{3}^{n}\\
\vdots&\vdots&\vdots&\vdots&\vdots&\vdots&\vdots\\
-{\gamma_{2n-1}}^{\ast}&1&{\lambda_{2n-1}}^{\ast}{\gamma_{2n-1}}^{\ast}&\ldots&-(-{\lambda_{2n-1}}^{\ast})^{n-1}{\gamma_{2n-1}}^{\ast}&(-{\lambda_{2n-1}}^{\ast})^{n-1}&-(-{\lambda_{2n-1}}^{\ast})^{n}{\gamma_{2n-1}}^{\ast}\nonumber\\
\end{vmatrix},
\end{equation*}
\begin{equation*}\label{fsst4}
\widehat{(T_{n})_{21}}=\begin{vmatrix}
1&0&\lambda&\ldots&\lambda^{n-1}&0&0\\
1&\gamma_{1}&\lambda_{1}&\ldots&\lambda_{1}^{n-1}&\lambda_{1}^{n-1}\gamma_{1}&\lambda_{1}^{n}\gamma_{1}\\
-{\gamma_{1}}^{\ast}&1&{\lambda_{1}}^{\ast}{\gamma_{1}}^{\ast}&\ldots&-(-{\lambda_{1}}^{\ast})^{n-1}{\gamma_{1}}^{\ast}&(-{\lambda_{1}}^{\ast})^{n-1}&(-{\lambda_{1}}^{\ast})^{n}\\
1&\gamma_{3}&\lambda_{3}&\ldots&\lambda_{3}^{n-1}&\lambda_{3}^{n-1}\gamma_{3}&\lambda_{3}^{n}\gamma_{3}\\
\vdots&\vdots&\vdots&\vdots&\vdots&\vdots&\vdots\\
-{\gamma_{2n-1}}^{\ast}&1&{\lambda_{2n-1}}^{\ast}{\gamma_{2n-1}}^{\ast}&\ldots&-(-{\lambda_{2n-1}}^{\ast})^{n-1}{\gamma_{2n-1}}^{\ast}&(-{\lambda_{2n-1}}^{\ast})^{n-1}&(-{\lambda_{2n-1}}^{\ast})^{n}\nonumber\\
\end{vmatrix},
\end{equation*}
\begin{equation*}\label{fsst5}
\widehat{(T_{n})_{22}}=\begin{vmatrix}
0&1&0&\ldots&0&\lambda^{n-1}&\lambda^{n}\\
1&\gamma_{1}&\lambda_{1}&\ldots&\lambda_{1}^{n-1}&\lambda_{1}^{n-1}\gamma_{1}&\lambda_{1}^{n}\gamma_{1}\\
-{\gamma_{1}}^{\ast}&1&{\lambda_{1}}^{\ast}{\gamma_{1}}^{\ast}&\ldots&-(-{\lambda_{1}}^{\ast})^{n-1}{\gamma_{1}}^{\ast}&(-{\lambda_{1}}^{\ast})^{n-1}&(-{\lambda_{1}}^{\ast})^{n}\\
1&\gamma_{3}&\lambda_{3}&\ldots&\lambda_{3}^{n-1}&\lambda_{3}^{n-1}\gamma_{3}&\lambda_{3}^{n}\gamma_{3}\\
\vdots&\vdots&\vdots&\vdots&\vdots&\vdots&\vdots\\
-{\gamma_{2n-1}}^{\ast}&1&{\lambda_{2n-1}}^{\ast}{\gamma_{2n-1}}^{\ast}&\ldots&-(-{\lambda_{2n-1}}^{\ast})^{n-1}{\gamma_{2n-1}}^{\ast}&(-{\lambda_{2n-1}}^{\ast})^{n-1}&(-{\lambda_{2n-1}}^{\ast})^{n}\nonumber\\
\end{vmatrix},
\end{equation*}
\begin{eqnarray*}
\mbox{\hspace{-0.8cm}}\widehat{(Q_{n})_{11}}=\begin{vmatrix}
1&\gamma_{1}&\lambda_{1}&\lambda_{1}\gamma_{1}&{\lambda_{1}}^2 &\ldots &{\lambda_{1}}^{n-1} \gamma_{1}&{\lambda_{1}}^{n}\\
-{\gamma_{1}}^{\ast}&1&{\lambda_{1}}^{\ast}{\gamma_{1}}^{\ast}&-{\lambda_{1}}^{\ast}&-{{\lambda_{1}}^{\ast}}^2{\gamma_{1}}^{\ast}&\ldots&{(-{\lambda_{1}}^{\ast})}^{n-1}&-{(-{\lambda_{1}}^{\ast}})^{n}{\gamma_{1}}^{\ast}\\
1&\gamma_{3}&\lambda_{3}&\lambda_{3}\gamma_{3}&{\lambda_{3}}^2 &\ldots &{\lambda_{3}}^{n-1} \gamma_{3}&{\lambda_{3}}^{n}\\
\vdots&\vdots&\vdots&\vdots&\vdots&\vdots&\vdots&\vdots\\
-{\gamma_{2n-1}}^{\ast}&1&{\lambda_{2n-1}}^{\ast}{\gamma_{2n-1}}^{\ast}&-{\lambda_{2n-1}}^{\ast}&-{{\lambda_{2n-1}}^{\ast}}^2{\gamma_{2n-1}}^{\ast}&\ldots&{(-{\lambda_{2n-1}}^{\ast})}^{n-1}&-{(-{\lambda_{2n-1}}^{\ast}})^{n}{\gamma_{2n-1}}^{\ast}\nonumber\\
\end{vmatrix},
\end{eqnarray*}
\begin{equation*}
\widehat{(Q_{n})_{12}}=\begin{vmatrix}
1&\gamma_{1}&\lambda_{1}&\lambda_{1}\gamma_{1} &\ldots&{\lambda_{1}}^{n-1} &{\lambda_{1}}^{n}\\
-{\gamma_{1}}^{\ast}&1&{\lambda_{1}}^{\ast}{\gamma_{1}}^{\ast}&-{\lambda_{1}}^{\ast}&\ldots&-{(-{\lambda_{1}}^{\ast}})^{n-1}{\gamma_{1}}^{\ast}&-{(-{\lambda_{1}}^{\ast}})^{n}{\gamma_{1}}^{\ast}\\
1&\gamma_{3}&\lambda_{3}&\lambda_{3}\gamma_{3}&\ldots&{\lambda_{3}}^{n-1} &{\lambda_{3}}^{n}\\
\vdots&\vdots&\vdots&\vdots&\vdots&\vdots&\vdots\\
-{\gamma_{2n-1}}^{\ast}&1&{\lambda_{2n-1}}^{\ast}{\gamma_{2n-1}}^{\ast}&-{\lambda_{2n-1}}^{\ast}&\ldots&-{(-{\lambda_{2n-1}}^{\ast}})^{n-1}{\gamma_{2n-1}}^{\ast}&-{(-{\lambda_{2n-1}}^{\ast}})^{n}{\gamma_{2n-1}}^{\ast}\nonumber\\
\end{vmatrix},
\end{equation*}
\begin{equation*}
\widehat{(Q_{n})_{21}}=\begin{vmatrix}
1&\gamma_{1}&\lambda_{1}&\lambda_{1}\gamma_{1}&{\lambda_{1}}^2 &\ldots&{\lambda_{1}}^{n-1} \gamma_{1}&{\lambda_{1}}^{n} \gamma_{1}\\
-{\gamma_{1}}^{\ast}&1&{\lambda_{1}}^{\ast}{\gamma_{1}}^{\ast}&-{\lambda_{1}}^{\ast}&-{{\lambda_{1}}^{\ast}}^2{\gamma_{1}}^{\ast}&\ldots&{(-{\lambda_{1}}^{\ast})}^{n-1}&{(-{\lambda_{1}}^{\ast})}^{n}\\
1&\gamma_{3}&\lambda_{3}&\lambda_{3}\gamma_{3}&{\lambda_{3}}^2 &\ldots&{\lambda_{3}}^{n-1} \gamma_{3} &{\lambda_{3}}^{n} \gamma_{3}\\
\vdots&\vdots&\vdots&\vdots&\vdots&\vdots&\vdots&\vdots\\
-{\gamma_{2n-1}}^{\ast}&1&{\lambda_{2n-1}}^{\ast}{\gamma_{2n-1}}^{\ast}&-{\lambda_{2n-1}}^{\ast}&-{{\lambda_{2n-1}}^{\ast}}^2{\gamma_{2n-1}}^{\ast}&\ldots&{(-{\lambda_{2n-1}}^{\ast})}^{n-1}&{(-{\lambda_{2n-1}}^{\ast})}^{n}\nonumber\\
\end{vmatrix},
\end{equation*}
\begin{equation*}
\mbox{\hspace{-0.8cm}}\widehat{(Q_{n})_{22}}=\begin{vmatrix}
1&\gamma_{1}&\lambda_{1}&\lambda_{1}\gamma_{1}&{\lambda_{1}}^2 &\ldots&{\lambda_{1}}^{n-1} &{\lambda_{1}}^{n} \gamma_{1}\\
-{\gamma_{1}}^{\ast}&1&{\lambda_{1}}^{\ast}{\gamma_{1}}^{\ast}&-{\lambda_{1}}^{\ast}&-{{\lambda_{1}}^{\ast}}^2{\gamma_{1}}^{\ast}&\ldots&-{(-{\lambda_{1}}^{\ast}})^{n-1}{\gamma_{1}}^{\ast}&{(-{\lambda_{1}}^{\ast})}^{n}\\
1&\gamma_{3}&\lambda_{3}&\lambda_{3}\gamma_{3}&{\lambda_{3}}^2 &\ldots&{\lambda_{3}}^{n-1} &{\lambda_{3}}^{n} \gamma_{3}\\
\vdots&\vdots&\vdots&\vdots&\vdots&\vdots&\vdots&\vdots\\
-{\gamma_{2n-1}}^{\ast}&1&{\lambda_{2n-1}}^{\ast}{\gamma_{2n-1}}^{\ast}&-{\lambda_{2n-1}}^{\ast}&-{{\lambda_{2n-1}}^{\ast}}^2{\gamma_{2n-1}}^{\ast}&\ldots&-{(-{\lambda_{2n-1}}^{\ast}})^{n-1}{\gamma_{2n-1}}^{\ast}&{(-{\lambda_{2n-1}}^{\ast})}^{n}\nonumber\\
\end{vmatrix}.
\end{equation*}
Here
\begin{eqnarray}
&&\gamma_{2k-1}=\dfrac{v_1}{v_2},\label{bi11}
\end{eqnarray}
\begin{eqnarray*}
&&v_1=\lefteqn{(2 i( \alpha_{2k-1}^2-  d^2)
\sin(\dfrac{\sqrt{d^2-{\alpha_{2k-1}}^2}(-s_2 x+s_4
t)}{s_2})\cos(\dfrac{\sqrt{d^2-{\alpha_{2k-1}}^2}(-s_2 x+s_4
t)}{s_2}){}}\nonumber\\&&{}+i(-{\alpha_{2k-1}}^2+d^2)\sin(\dfrac{2\sqrt{d^2-{\alpha_{2k-1}}^2}(-s_2
x+s_4 t)}{s2})-2 d \alpha_{2k-1}
\cosh(\dfrac{2\sqrt{d^2-{\alpha_{2k-1}}^2} s_3
t}{s_2})\nonumber\\&&{}+2 d^2
\cos(\dfrac{2\sqrt{d^2-{\alpha_{2k-1}}^2}(-s_2 x+s_4 t)}{s_2})+2 i d
\sinh(\dfrac{2\sqrt{d^2-{\alpha_{2k-1}}^2} s_3 t}{s_2})\sqrt{d^2-{\alpha_{2k-1}}^2})\exp(-i s_1),\nonumber\\
&&v_2=\lefteqn{(-2 d^2+2 {\alpha_{2k-1}}^2)\sinh(\dfrac{\sqrt{d^2-{\alpha_{2k-1}}^2}  s_3 t}{s_2}) \cosh(\dfrac{\sqrt{d^2-{\alpha_{2k-1}}^2} s_3 t }{s_2})+2 d^2 \cosh(\dfrac{2\sqrt{d^2-{\alpha_{2k-1}}^2} s_3 t }{s_2}){}}\nonumber\\&&{}+(d^2-{\alpha_{2k-1}}^2) \sinh(\dfrac{2\sqrt{d^2-{\alpha_{2k-1}}^2}  s_3 t}{s_2})-2 \sqrt{d^2-{\alpha_{2k-1}}^2} d \sin(\dfrac{2\sqrt{d^2-{\alpha_{2k-1}}^2}(-s_2 x+s_4 t)}{s_2})\nonumber\\&&{}-2 d \alpha_{2k-1} \cos(2\dfrac{\sqrt{d^2-{\alpha_{2k-1}}^2}(-s_2 x+s_4 t)}{s_2}),\nonumber\\
&&s_1= \dfrac{-b(1/2b-\omega_0)x+(2+2(1/2b-\omega_0)(1/4
b^2-1/2d^2))t}{-1
/2b+\omega_0},\nonumber\\
&&s_2= (-1/2 b+\omega_0)((1/2 b-\omega_0)^2+d^2),\nonumber\\
&&s_3= -d \omega_0 (d^2+\omega_0^2)-3/2 d b\omega_0 (1/2
b-\omega_0)+1/2 d b
(d^2+1/4 b^2)+d,\nonumber\\
&&s_4= -1/2 b^2 (d^2+1/4 b^2)+b \omega_0(d^2+{\omega_0}^2)+3/2 b^2
\omega_0(1
/2b-\omega_0)+1/2 b-\omega_0.\nonumber\\
\end{eqnarray*}
\newpage
\begin{center}
\appendix{\textbf{Appendix III: Here,
we are giving the construction of T of
 $\bar{T}_{rn}(\lambda)$} in detail.}
\end{center}
\setcounter{equation}{0}
\renewcommand{\theequation}{B.\arabic{equation}}
\begin{equation}\label{rfss1}\bar{T}_{rn}=\bar{T}_{rn}(\lambda)=\left(
\begin{array}{cc}
\dfrac{\widetilde{(T_{rn})_{11}}}{|W_{r2n}|}& \dfrac{\widetilde{(T_{rn})_{12}}}{|W_{r2n}|}\\ \\
\dfrac{\widetilde{(T_{rn})_{21}}}{|W_{r2n}|}& \dfrac{\widetilde{(T_{rn})_{22}}}{|W_{r2n}|}\\
\end{array} \right),
\end{equation}

\begin{eqnarray*}
\bar{t}_{r1}=\left(
\begin{array}{cc}
\dfrac{\widetilde{(Q_{rn})_{11}}}{|W_{r2n}|}& \dfrac{\widetilde{(Q_{rn})_{12}}}{|W_{r2n}|}\\ \\
\dfrac{\widetilde{(Q_{rn})_{21}}}{|W_{r2n}|}& \dfrac{\widetilde{(Q_{rn})_{22}}}{|W_{r2n}|}\\
\end{array} \right),
\end{eqnarray*}

\begin{equation*}\label{fsst1}
W_{r2n}=\left(\begin{matrix}
h_{01}^{1}&h_{02}^{1}&h_{11}^{1}&h_{12}^{1}& \ldots&h_{n-11}^{1}&h_{n-12}^{1}\\
-{h_{02}^{1}}^{\ast}&{h_{01}^{1}}^{\ast}&{h_{12}^{1}}^{\ast}&-{h_{11}^{1}}^{\ast}&\ldots&(-1)^{n}{h_{n-12}^{1}}^{\ast}&(-1)^{n-1}{h_{n-11}^{1}}^{\ast}\\
h_{01}^{3}&h_{02}^{3}&h_{11}^{3}&h_{12}^{3}&\ldots&h_{n-11}^{3}&h_{n-12}^{3}\\
\vdots&\vdots&\vdots&\vdots&\vdots&\vdots&\vdots\\
-{h_{02}^{2n-1}}^{\ast}&{h_{01}^{2n-1}}^{\ast}&{h_{12}^{2n-1}}^{\ast}&-{h_{11}^{2n-1}}^{\ast}&\ldots&(-1)^{n}{h_{n-12}^{2n-1}}^{\ast}&(-1)^{n-1}{h_{n-11}^{2n-1}}^{\ast}\nonumber\\
\end{matrix}\right),
\end{equation*}
\begin{equation*}\label{fsst2}
\widetilde{(T_{rn})_{11}}=\begin{vmatrix}
1&0&\lambda&0&\ldots&\lambda^{n-1}&0&\lambda^{n}\\
h_{01}^{1}&h_{02}^{1}&h_{11}^{1}&h_{12}^{1}&\ldots&h_{n-11}^{1}&h_{n-12}^{1}&h_{n1}^{1}\\
-{h_{02}^{1}}^{\ast}&{h_{01}^{1}}^{\ast}&{h_{12}^{1}}^{\ast}&-{h_{11}^{1}}^{\ast}&\ldots&(-1)^{n}{h_{n-12}^{1}}^{\ast}&(-1)^{n-1}{h_{n-11}^{1}}^{\ast}&(-1)^{n+1}{{h_{n2}}^{1}}^{\ast}\\
h_{01}^{3}&h_{02}^{3}&h_{11}^{3}&h_{12}^{3}&\ldots&h_{n-11}^{3}&h_{n-12}^{3}&h_{n1}^{3}\\
\vdots&\vdots&\vdots&\vdots&\vdots&\vdots&\vdots&\vdots\\
-{h_{02}^{2n-1}}^{\ast}&{h_{01}^{2n-1}}^{\ast}&{h_{12}^{2n-1}}^{\ast}&-{h_{11}^{2n-1}}^{\ast}&\ldots&(-1)^{n}{h_{n-12}^{2n-1}}^{\ast}&(-1)^{n-1}{h_{n-11}^{2n-1}}^{\ast}&(-1)^{n+1}{{h_{n2}}^{2n-1}}^{\ast}\nonumber\\
\end{vmatrix},
\end{equation*}
\begin{equation*}\label{fsst3}
\widetilde{(T_{rn})_{12}}=\begin{vmatrix}
0&1&0&\lambda&\ldots&0&\lambda^{n-1}&0\\
h_{01}^{1}&h_{02}^{1}&h_{11}^{1}&h_{12}^{1}&\ldots&h_{n-11}^{1}&h_{n-12}^{1}&h_{n1}^{1}\\
-{h_{02}^{1}}^{\ast}&{h_{01}^{1}}^{\ast}&{h_{12}^{1}}^{\ast}&-{h_{11}^{1}}^{\ast}&\ldots&(-1)^{n}{h_{n-12}^{1}}^{\ast}&(-1)^{n-1}{h_{n-11}^{1}}^{\ast}&(-1)^{n+1}{{h_{n2}}^{1}}^{\ast}\\
h_{01}^{3}&h_{02}^{3}&h_{11}^{3}&h_{12}^{3}&\ldots&h_{n-11}^{3}&h_{n-12}^{3}&h_{n1}^{3}\\
\vdots&\vdots&\vdots&\vdots&\vdots&\vdots&\vdots&\vdots\\
-{h_{02}^{2n-1}}^{\ast}&{h_{01}^{2n-1}}^{\ast}&{h_{12}^{2n-1}}^{\ast}&-{h_{11}^{2n-1}}^{\ast}&\ldots&(-1)^{n}{h_{n-12}^{2n-1}}^{\ast}&(-1)^{n-1}{h_{n-11}^{2n-1}}^{\ast}&(-1)^{n+1}{{h_{n2}}^{2n-1}}^{\ast}\nonumber\\
\end{vmatrix},
\end{equation*}
\begin{equation*}\label{fsst4}
\widetilde{(T_{rn})_{21}}=\begin{vmatrix}
1&0&\lambda&0&\ldots&\lambda^{n-1}&0&0\\
h_{01}^{1}&h_{02}^{1}&h_{11}^{1}&h_{12}^{1}&\ldots&h_{n-11}^{1}&h_{n-12}^{1}&h_{n2}^{1}\\
-{h_{02}^{1}}^{\ast}&{h_{01}^{1}}^{\ast}&{h_{12}^{1}}^{\ast}&-{h_{11}^{1}}^{\ast}&\ldots&(-1)^{n}{h_{n-12}^{1}}^{\ast}&(-1)^{n-1}{h_{n-11}^{1}}^{\ast}&(-1)^{n}{h_{n1}^{1}}^{\ast}\\
h_{01}^{3}&h_{02}^{3}&h_{11}^{3}&h_{12}^{3}&\ldots&h_{n-11}^{3}&h_{n-12}^{3}&h_{n2}^{3}\\
\vdots&\vdots&\vdots&\vdots&\vdots&\vdots&\vdots&\vdots\\
-{h_{02}^{2n-1}}^{\ast}&{h_{01}^{2n-1}}^{\ast}&{h_{12}^{2n-1}}^{\ast}&-{h_{11}^{2n-1}}^{\ast}&\ldots&(-1)^{n}{h_{n-12}^{2n-1}}^{\ast}&(-1)^{n-1}{h_{n-11}^{2n-1}}^{\ast}&(-1)^{n}{h_{n1}^{2n-1}}^{\ast}\nonumber\\
\end{vmatrix},
\end{equation*}
\begin{equation*}\label{fsst5}
\widetilde{(T_{rn})_{22}}=\begin{vmatrix}
0&1&0&\lambda&\ldots&0&\lambda^{n-1}&\lambda^{n}\\
h_{01}^{1}&h_{02}^{1}&h_{11}^{1}&h_{12}^{1}&\ldots&h_{n-11}^{1}&h_{n-12}^{1}&h_{n2}^{1}\\
-{h_{02}^{1}}^{\ast}&{h_{01}^{1}}^{\ast}&{h_{12}^{1}}^{\ast}&-{h_{11}^{1}}^{\ast}&\ldots&(-1)^{n}{h_{n-12}^{1}}^{\ast}&(-1)^{n-1}{h_{n-11}^{1}}^{\ast}&(-1)^{n}{h_{n1}^{1}}^{\ast}\\
h_{01}^{3}&h_{02}^{3}&h_{11}^{3}&h_{12}^{3}&\ldots&h_{n-11}^{3}&h_{n-12}^{3}&h_{n2}^{3}\\
\vdots&\vdots&\vdots&\vdots&\vdots&\vdots&\vdots&\vdots\\
-{h_{02}^{2n-1}}^{\ast}&{h_{01}^{2n-1}}^{\ast}&{h_{12}^{2n-1}}^{\ast}&-{h_{11}^{2n-1}}^{\ast}&\ldots&(-1)^{n}{h_{n-12}^{2n-1}}^{\ast}&(-1)^{n-1}{h_{n-11}^{2n-1}}^{\ast}&(-1)^{n}{h_{n1}^{2n-1}}^{\ast}\nonumber\\
\end{vmatrix},
\end{equation*}
\begin{equation*}
\widetilde{(Q_{rn})_{11}}=\begin{vmatrix}
h_{01}^{1}&h_{02}^{1}&h_{11}^{1}&h_{12}^{1}&\ldots&h_{n-12}^{1}&h_{n1}^{1}\\
-{h_{02}^{1}}^{\ast}&{h_{01}^{1}}^{\ast}&{h_{12}^{1}}^{\ast}&-{h_{11}^{1}}^{\ast}&\ldots&(-1)^{n-1}{h_{n-11}^{1}}^{\ast}&(-1)^{n+1}{{h_{n2}}^{1}}^{\ast}\\
h_{01}^{3}&h_{02}^{3}&h_{11}^{3}&h_{12}^{3}&\ldots&h_{n-12}^{3}&h_{n1}^{3}\\
\vdots&\vdots&\vdots&\vdots&\vdots&\vdots&\vdots\\
-{h_{02}^{2n-1}}^{\ast}&{h_{01}^{2n-1}}^{\ast}&{h_{12}^{2n-1}}^{\ast}&-{h_{11}^{2n-1}}^{\ast}&\ldots&(-1)^{n-1}{h_{n-11}^{2n-1}}^{\ast}&(-1)^{n+1}{{h_{n2}}^{2n-1}}^{\ast}\nonumber\\
\end{vmatrix},
\end{equation*}
\begin{equation*}
\widetilde{(Q_{rn})_{12}}=-\begin{vmatrix}
h_{01}^{1}&h_{02}^{1}&h_{11}^{1}&h_{12}^{1}&\ldots&h_{n-11}^{1}&h_{n1}^{1}\\
-{h_{02}^{1}}^{\ast}&{h_{01}^{1}}^{\ast}&{h_{12}^{1}}^{\ast}&-{h_{11}^{1}}^{\ast}&\ldots&(-1)^{n}{h_{n-12}^{1}}^{\ast}&(-1)^{n+1}{{h_{n2}}^{1}}^{\ast}\\
h_{01}^{3}&h_{02}^{3}&h_{11}^{3}&h_{12}^{3}&\ldots&h_{n-11}^{3}&h_{n1}^{3}\\
\vdots&\vdots&\vdots&\vdots&\vdots&\vdots&\vdots\\
-{h_{02}^{2n-1}}^{\ast}&{h_{01}^{2n-1}}^{\ast}&{h_{12}^{2n-1}}^{\ast}&-{h_{11}^{2n-1}}^{\ast}&\ldots&(-1)^{n}{h_{n-12}^{2n-1}}^{\ast}&(-1)^{n+1}{{h_{n2}}^{2n-1}}^{\ast}\nonumber\\
\end{vmatrix},
\end{equation*}
\begin{equation*}
\widetilde{(Q_{rn})_{21}}=\begin{vmatrix}
h_{01}^{1}&h_{02}^{1}&h_{11}^{1}&h_{12}^{1}&\ldots&h_{n-12}^{1}&h_{n2}^{1}\\
-{h_{02}^{1}}^{\ast}&{h_{01}^{1}}^{\ast}&{h_{12}^{1}}^{\ast}&-{h_{11}^{1}}^{\ast}&\ldots&(-1)^{n-1}{h_{n-11}^{1}}^{\ast}&(-1)^{n}{h_{n1}^{1}}^{\ast}\\
h_{01}^{3}&h_{02}^{3}&h_{11}^{3}&h_{12}^{3}&\ldots&h_{n-12}^{3}&h_{n2}^{3}\\
\vdots&\vdots&\vdots&\vdots&\vdots&\vdots&\vdots\\
-{h_{02}^{2n-1}}^{\ast}&{h_{01}^{2n-1}}^{\ast}&{h_{12}^{2n-1}}^{\ast}&-{h_{11}^{2n-1}}^{\ast}&\ldots&(-1)^{n-1}{h_{n-11}^{2n-1}}^{\ast}&(-1)^{n}{h_{n1}^{2n-1}}^{\ast}\nonumber\\
\end{vmatrix},
\end{equation*}
\begin{equation*}
\widetilde{(Q_{rn})_{22}}=-\begin{vmatrix}
h_{01}^{1}&h_{02}^{1}&h_{11}^{1}&h_{12}^{1}&\ldots&h_{n-11}^{1}&h_{n2}^{1}\\
-{h_{02}^{1}}^{\ast}&{h_{01}^{1}}^{\ast}&{h_{12}^{1}}^{\ast}&-{h_{11}^{1}}^{\ast}&\ldots&(-1)^{n}{h_{n-12}^{1}}^{\ast}&(-1)^{n}{h_{n1}^{1}}^{\ast}\\
h_{01}^{3}&h_{02}^{3}&h_{11}^{3}&h_{12}^{3}&\ldots&h_{n-11}^{3}&h_{n2}^{3}\\
\vdots&\vdots&\vdots&\vdots&\vdots&\vdots&\vdots\\
-{h_{02}^{2n-1}}^{\ast}&{h_{01}^{2n-1}}^{\ast}&{h_{12}^{2n-1}}^{\ast}&-{h_{11}^{2n-1}}^{\ast}&\ldots&(-1)^{n}{h_{n-12}^{2n-1}}^{\ast}&(-1)^{n}{h_{n1}^{2n-1}}^{\ast}\nonumber\\
\end{vmatrix}.
\end{equation*}
Here
\begin{eqnarray*}
h_{mj}^{l}=\dfrac{\partial^{l}}{\partial{\delta}^{l}}((d+i\dfrac{b}{2}+\delta^2)^{m}f_{1j}(\lambda_1=d+i\dfrac{b}{2}+\delta^2))|_{\delta=0},
m=0,1,2,\cdots n,j=1,2,l=1,2,\cdots 2n.
\end{eqnarray*}
\end{widetext}


\begin{figure}[ht]
\setlength{\unitlength}{0.1cm}
\begin{minipage}[t]{8cm}
\epsfig{file=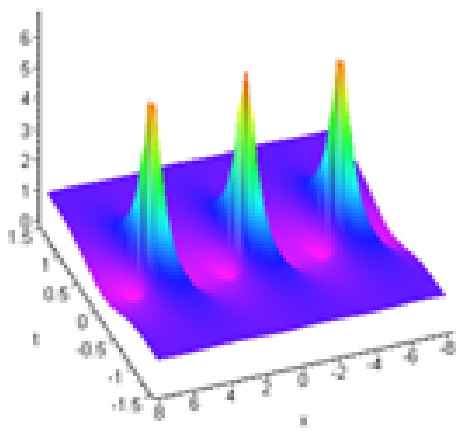,width=7cm}\vspace{-0.5cm}\caption{{ The
first order breather $|E^{[1]}|^{2}$ given  with specific parameters
$d=1,b=2,\omega_{0}=\dfrac{1}{2},\alpha_{1}=\dfrac{4}{5}$. There are
one upper peak and two caves in each periodic unit. }}
\end{minipage}
\hspace{0.5cm} 
\begin{minipage}[t]{8cm}
\epsfig{file=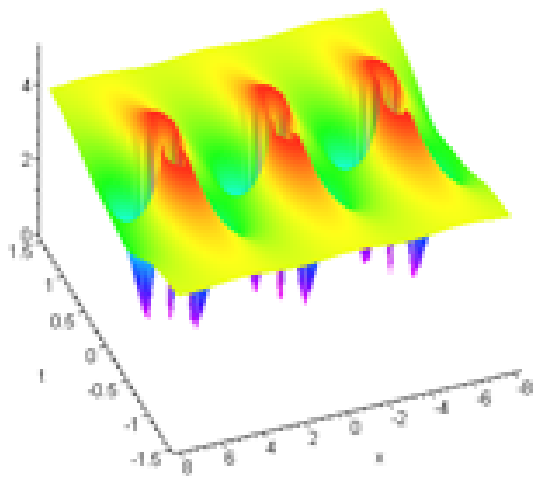,width=7cm}\vspace{-0.5cm}\caption{{The
first order dark breather $|p^{[1]}|^{2}$ for the values used in
Figure 1. There are a upper ring and three down peaks in each
periodic unit.}}
\end{minipage}
\end{figure}
\begin{figure}[ht]
\setlength{\unitlength}{0.1cm}
\begin{minipage}[t]{8cm}
\epsfig{file=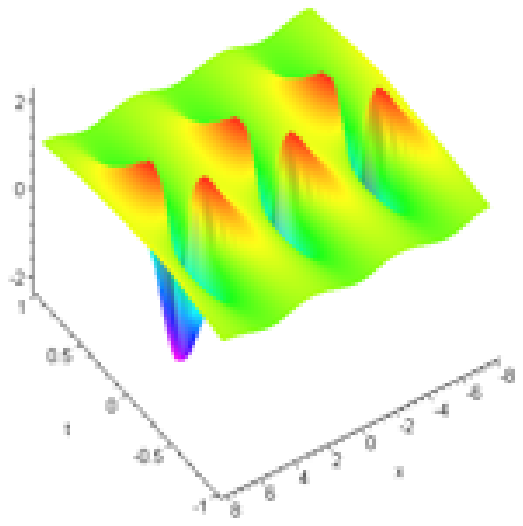,width=7cm}\vspace{-0.5cm}\caption{{The
first order dark breather $\eta^{[1]}$ for the values used in Figure
1. There are two lumps and one down peak in each periodic unit.}}
\end{minipage}
\hspace{0.5cm} 
\begin{minipage}[t]{8cm}
\epsfig{file=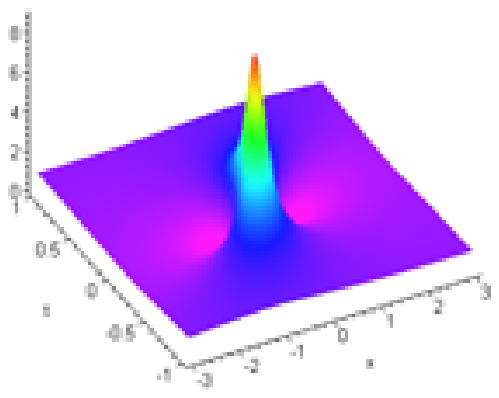,width=7cm}\vspace{-0.5cm}\caption{{The first
order rogue wave $|\tilde {E}^{[1]}|^{2}$ with specific parameters
$d=1,b=2,\omega_{0}=\dfrac{1}{2}$. There are one upper peak and two
caves.}}
\end{minipage}
\end{figure}
\begin{figure}[ht]
\setlength{\unitlength}{0.1cm}
\begin{minipage}[t]{8cm}
\epsfig{file=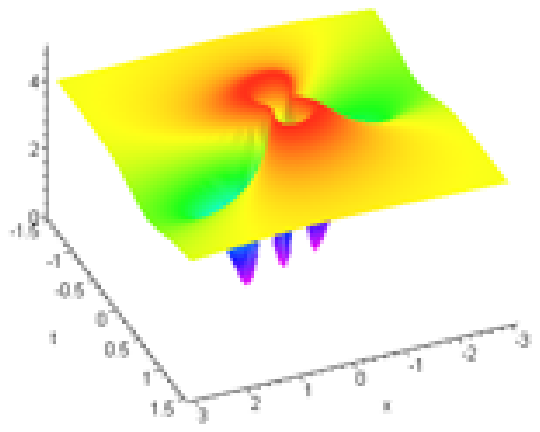,width=7cm}\vspace{-0.5cm}\caption{{The first
order dark rogue wave $|\tilde {p}^{[1]}|^{2}$  for the values used
in Figure 4. There are one upper ring and three down peaks.}}
\end{minipage}
\hspace{0.5cm} 
\begin{minipage}[t]{8cm}
\epsfig{file=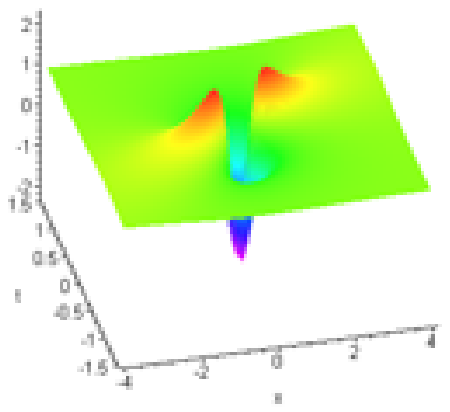,width=7cm}\vspace{-0.5cm}\caption{{The first
order dark rogue wave $\tilde {\eta}^{[1]}$  for the values used in
Figure 4. There are two lumps and one down peak.}}
\end{minipage}
\end{figure}

\begin{figure}[ht]
\setlength{\unitlength}{0.1cm}
\begin{minipage}[t]{8cm}
\epsfig{file=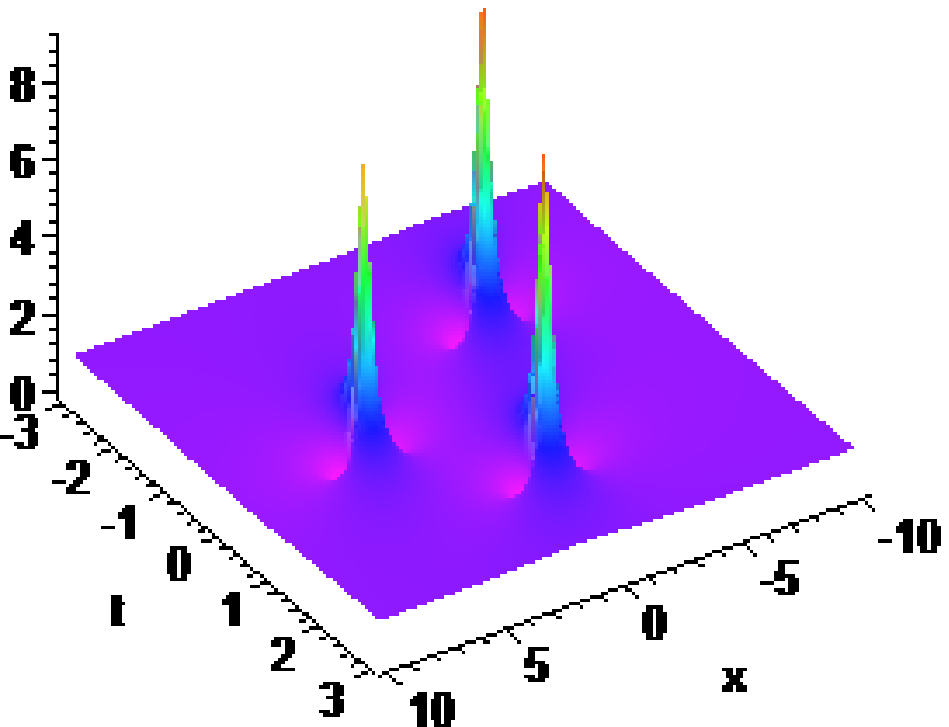,width=4cm}\vspace{0.2cm}\caption{{The second
order rogue wave $|\bar {E}^{[2]}|^{2}$ given by
eq.(\ref{rNLS-MBjien1})with specific parameters $d=1,b=2,
\omega_{0}=\dfrac{1}{2},K_0=1,J_0=0,J_1=100 $. }}
\end{minipage}
\hspace{0.5cm} 
\begin{minipage}[t]{8cm}
\hspace{0.5cm}\epsfig{file=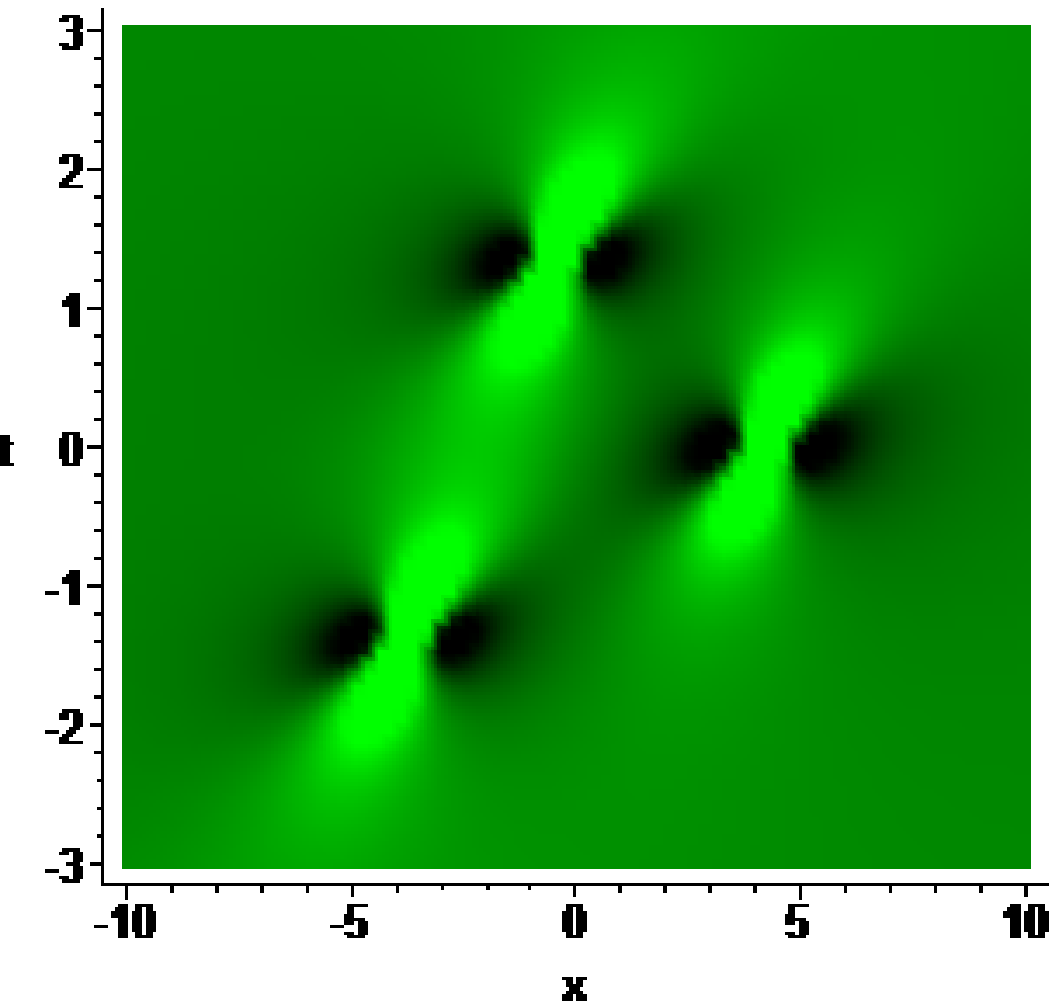,width=3cm}\vspace{0.2cm}\caption{{Contour
plot of the wave amplitudes of $|\bar {E}^{[2]}|^{2}$ for the values
used in Figure 7. }}
\end{minipage}
\end{figure}
\begin{figure}[ht]
\setlength{\unitlength}{0.1cm}
\begin{minipage}[t]{8cm}
\epsfig{file=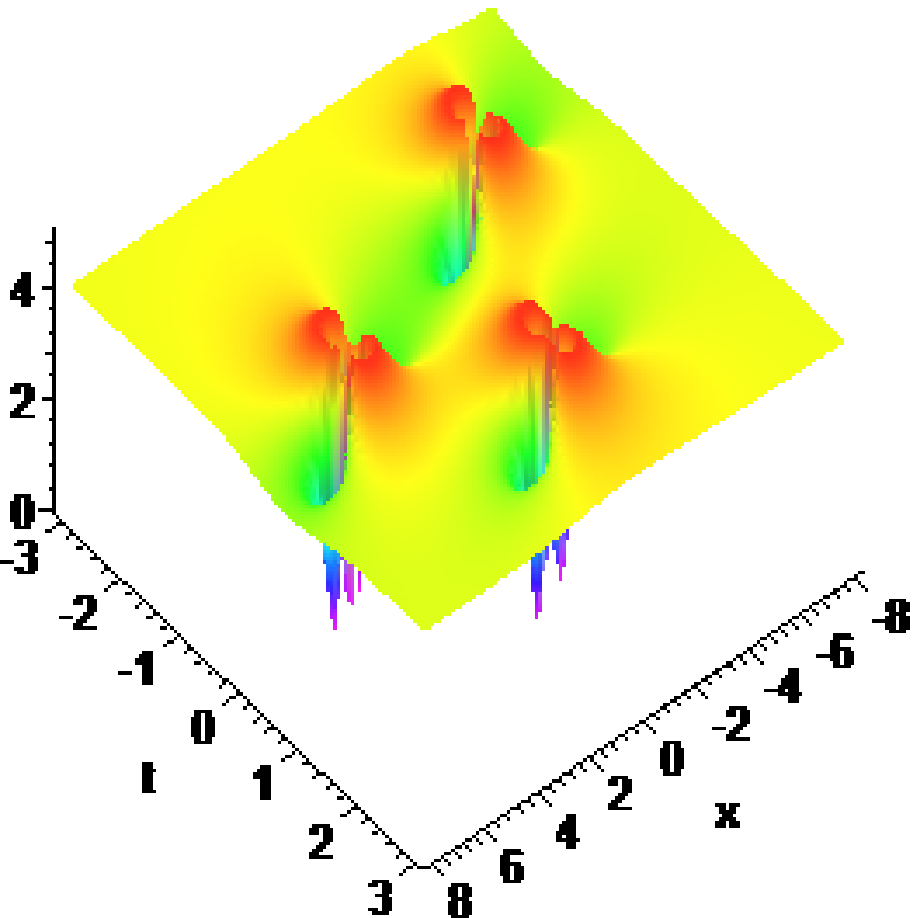,width=4cm}\vspace{0.2cm}\caption{{The second
order dark rogue wave $|\bar {p}^{[2]}|^{2}$ given by
eq.(\ref{rNLS-MBjien2}) for the values used in Figure 7.}}
\end{minipage}
\hspace{0.5cm} 
\begin{minipage}[t]{8cm}
\hspace{0.5cm}\epsfig{file=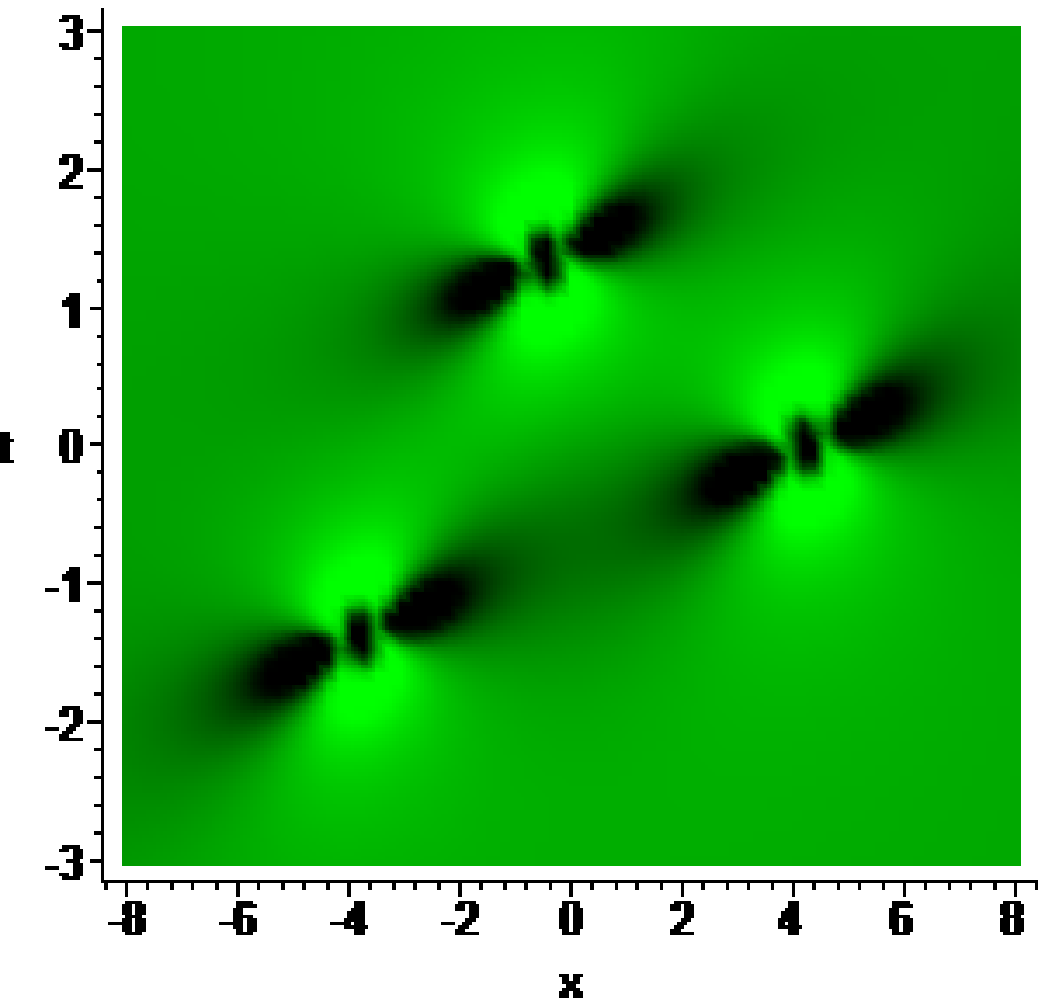,width=3cm}\vspace{0.2cm}\caption{{
Contour plot of the wave amplitudes of $|\bar {p}^{[2]}|^{2}$  for
the values used in Figure 7.}}
\end{minipage}
\end{figure}
\begin{figure}[ht]
\setlength{\unitlength}{0.1cm}
\begin{minipage}[t]{8cm}
\epsfig{file=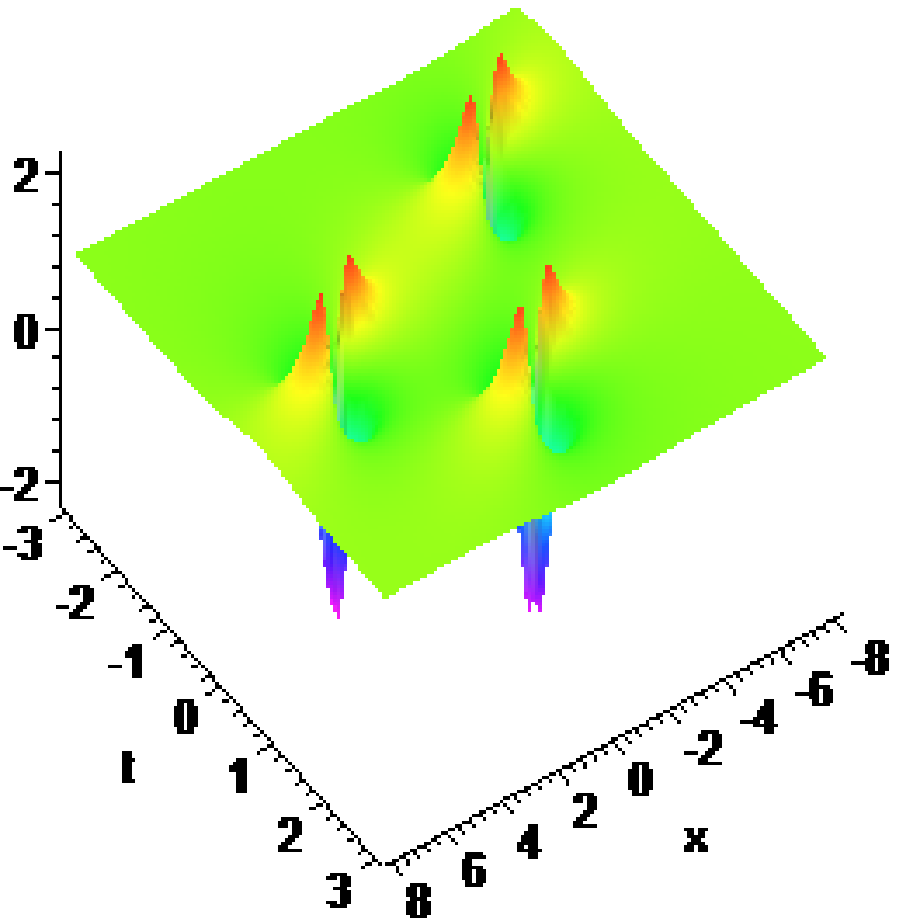,width=4cm}\vspace{0.2cm}\caption{{The second
order dark rogue wave $\bar {\eta}^{[2]} $ given by
eq.(\ref{rNLS-MBjien3}) for the values used in Figure 7. }}
\end{minipage}
\hspace{0.5cm} 
\begin{minipage}[t]{8cm}
\hspace{0.5cm}\epsfig{file=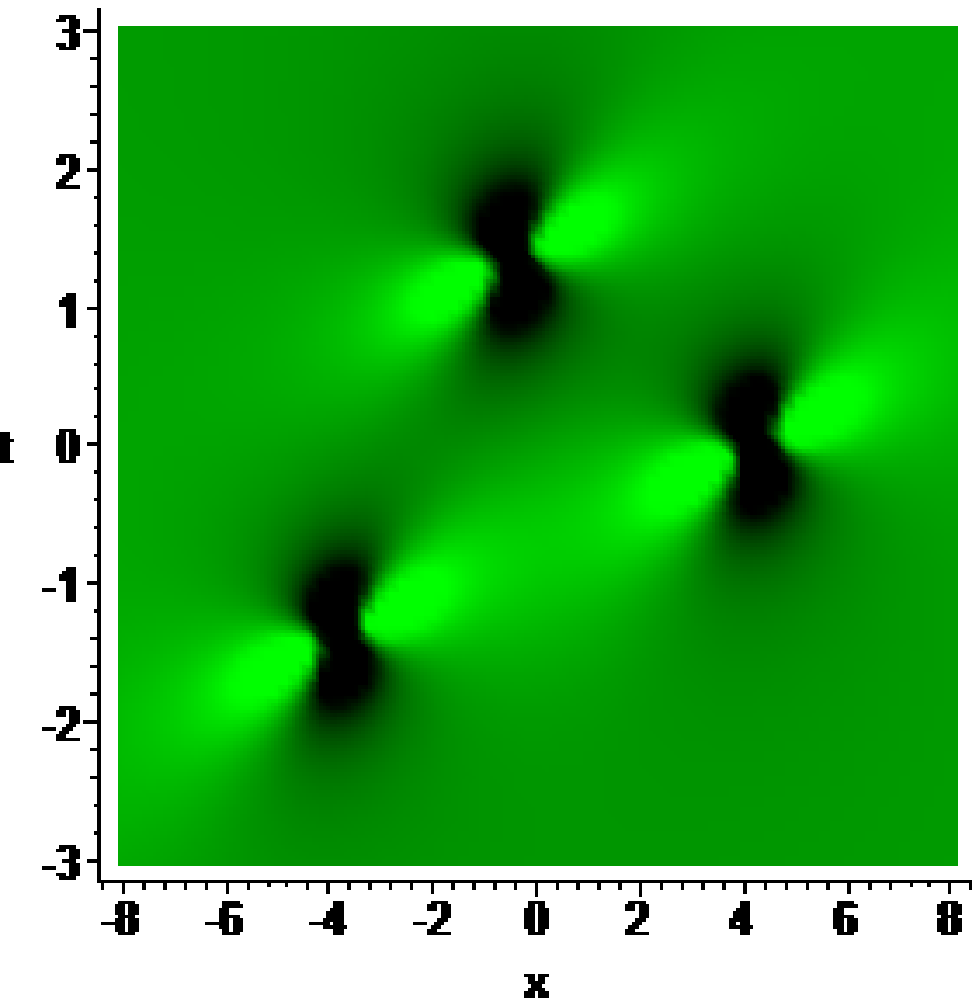,width=3cm}\vspace{0.2cm}\caption{{Contour
plot of the wave amplitudes of $\bar {\eta}^{[2]} $ for the values
used in Figure 7.}}
\end{minipage}
\end{figure}

\begin{figure}[ht]
\setlength{\unitlength}{0.1cm}
\begin{minipage}[t]{8cm}
\epsfig{file=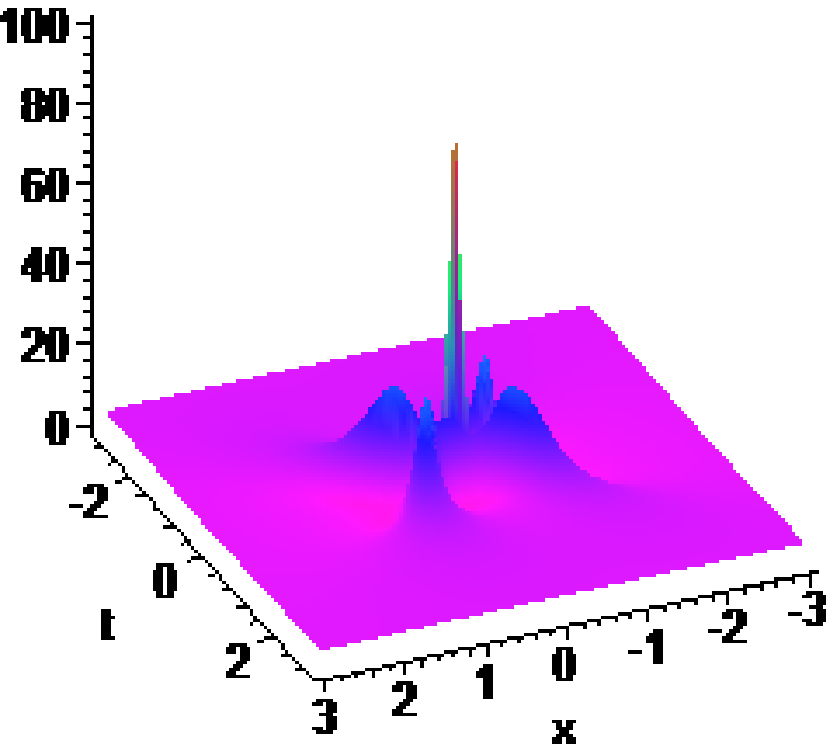,width=4cm}\vspace{0.2cm}\caption{{The second
order rogue wave $|\bar {E}^{[2]}|^{2}$ given by
eq.(\ref{rNLS-MBjien1})with specific parameters $d=2,b=0,
\omega_{0}=\dfrac{1}{2},K_0=1,J_0=0,J_1=0 $. }}
\end{minipage}
\hspace{0.5cm} 
\begin{minipage}[t]{8cm}
\epsfig{file=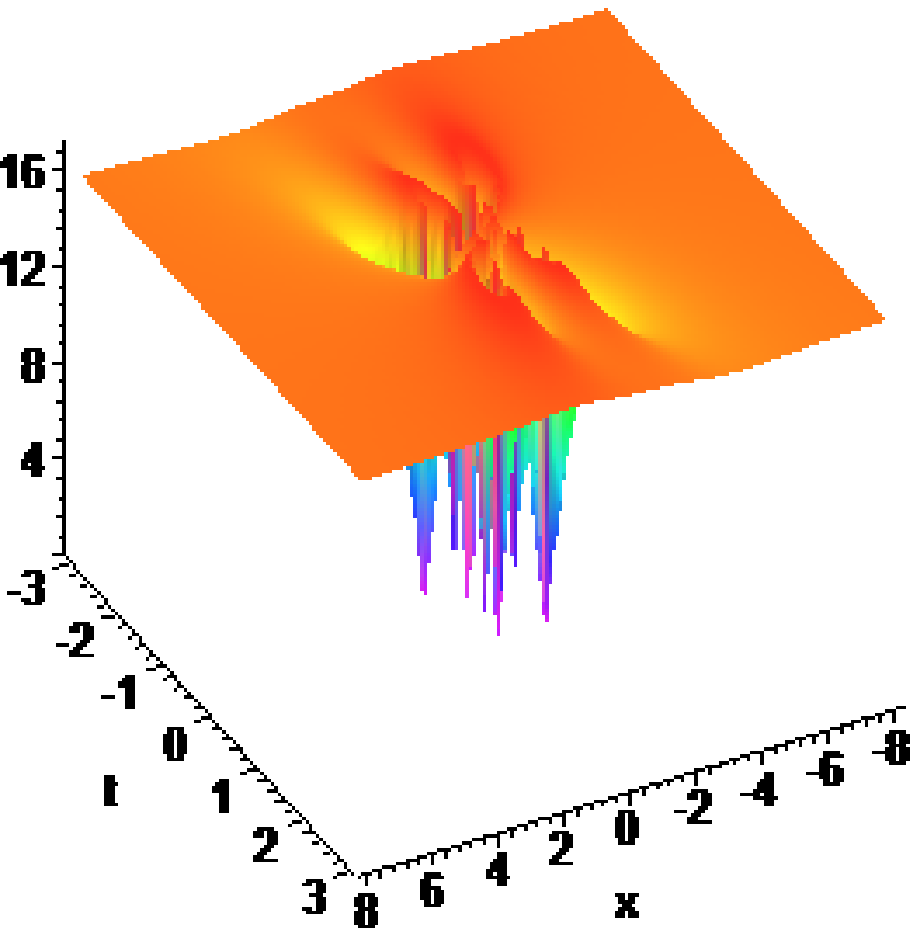,width=4cm}\vspace{0.2cm}\caption{{The second
order dark rogue wave $|\bar {p}^{[2]}|^{2}$ given by
eq.(\ref{rNLS-MBjien2}) for the values used in Figure 13.}}
\end{minipage}
\end{figure}

\begin{figure}[ht]
\setlength{\unitlength}{0.1cm}
\begin{minipage}[t]{8cm}
\epsfig{file=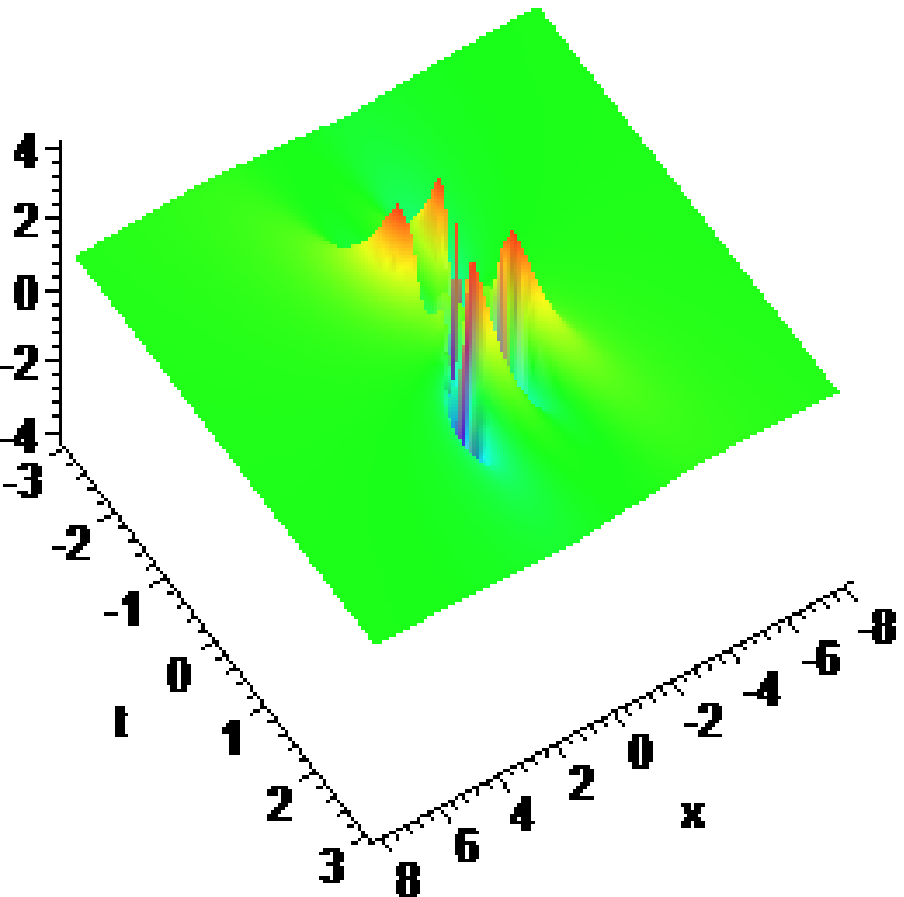,width=4cm}\vspace{0.2cm}\caption{{The second
order dark rogue wave $\bar {\eta}^{[2]} $ given by
eq.(\ref{rNLS-MBjien3}) for the values used in Figure 13. }}
\end{minipage}
\hspace{0.5cm} 
\begin{minipage}[t]{8cm}
\epsfig{file=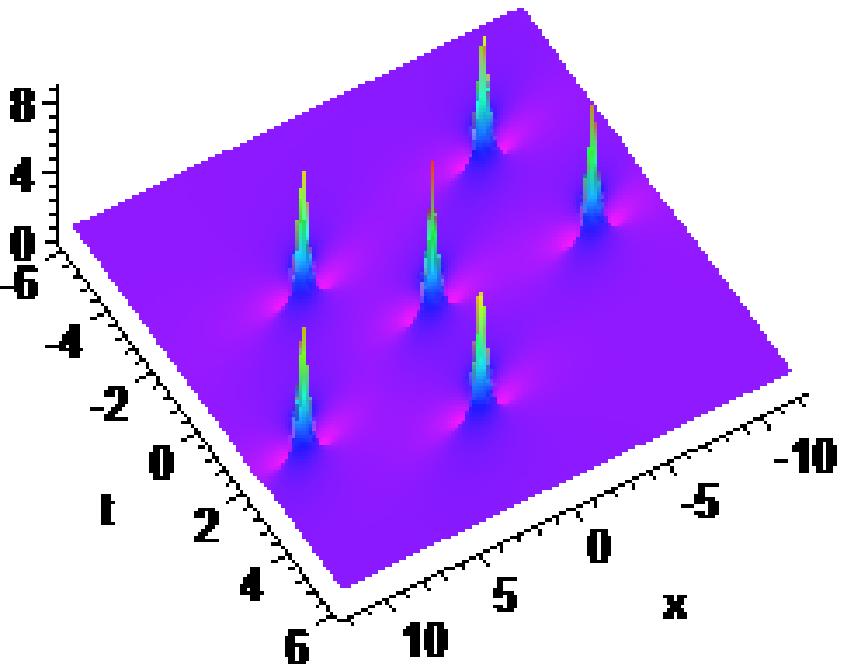,width=4cm}\vspace{0.2cm}\caption{{The third
order rogue wave $|\bar {E}^{[3]}|^{2}$ given by
eq.(\ref{rNLS-MBjien1})with specific parameters $d=1,b=2,
\omega_{0}=\dfrac{1}{2},K_0=1,J_0=0,J_1=0, J_2=8000 $. }}
\end{minipage}
\end{figure}
\begin{figure}[ht]
\setlength{\unitlength}{0.1cm}
\begin{minipage}[t]{8cm}
\epsfig{file=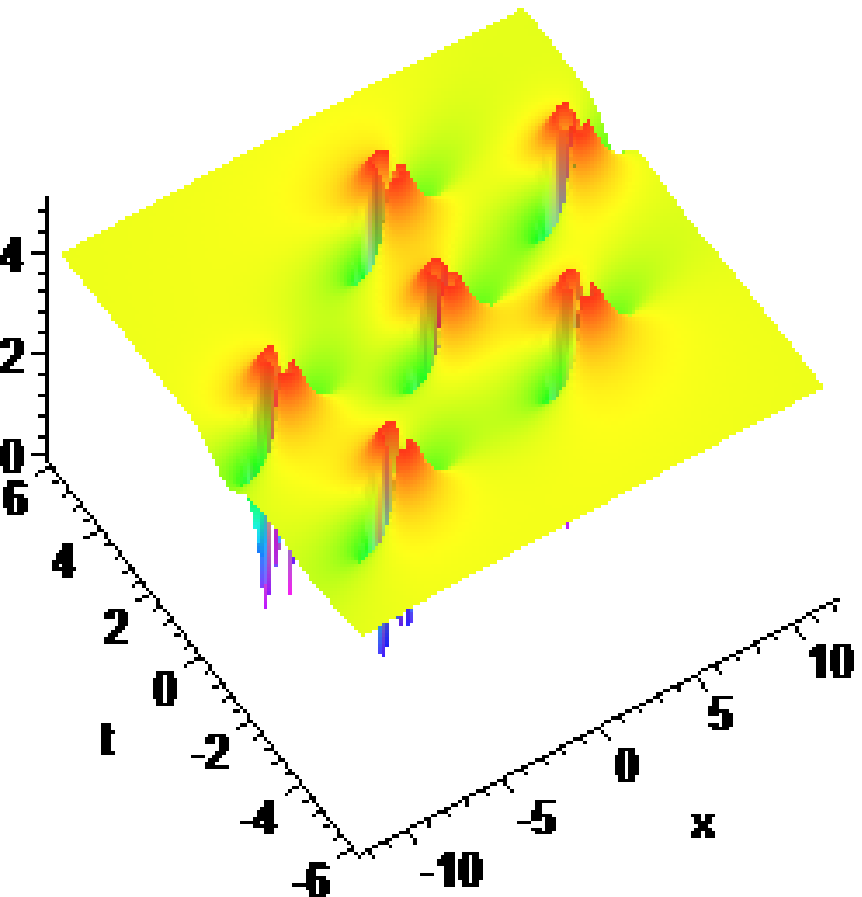,width=4cm}\vspace{0.2cm}\caption{{The third
order dark rogue wave $|\bar {p}^{[3]}|^{2}$ given by
eq.(\ref{rNLS-MBjien2}) for the values used in Figure 16.}}
\end{minipage}
\hspace{0.5cm} 
\begin{minipage}[t]{8cm}
\epsfig{file=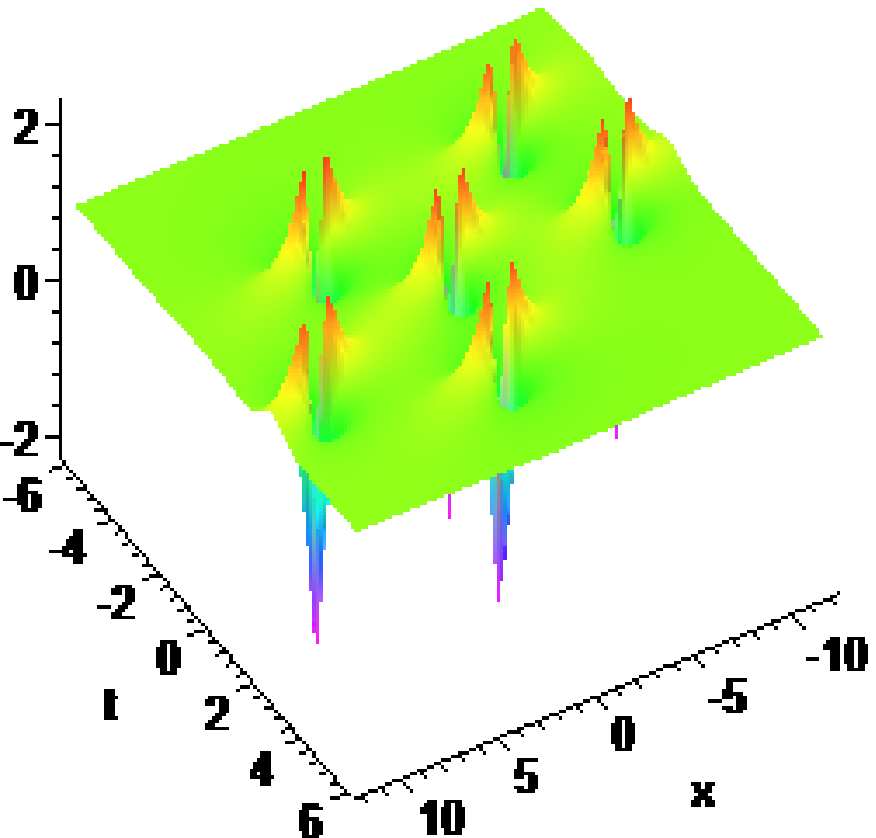,width=4cm}\vspace{0.2cm}\caption{{The third
order dark rogue wave $\bar {\eta}^{[3]} $ given by
eq.(\ref{rNLS-MBjien3}) for the values used in Figure 16. }}
\end{minipage}
\end{figure}


\begin{figure}[ht]
\setlength{\unitlength}{0.1cm}
\begin{minipage}[t]{8cm}
\epsfig{file=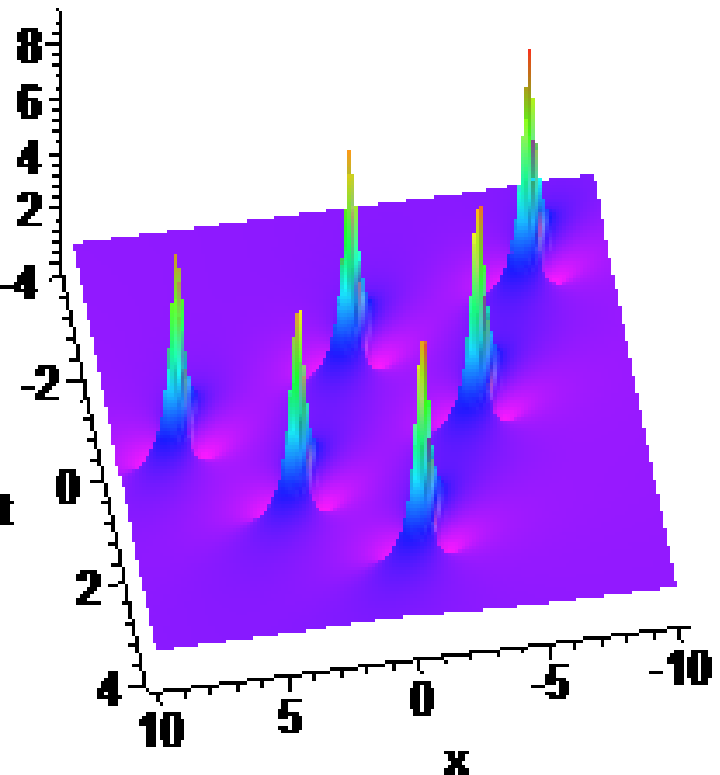,width=4cm}\vspace{0.2cm}\caption{{The third
order rogue wave $|\bar {E}^{[3]}|^{2}$ given by
eq.(\ref{rNLS-MBjien1})with specific parameters $d=1,b=2,
\omega_{0}=\dfrac{1}{2},K_0=1,J_0=0,J_1=100, J_2=0 $. }}
\end{minipage}
\hspace{0.5cm} 
\begin{minipage}[t]{8cm}
\epsfig{file=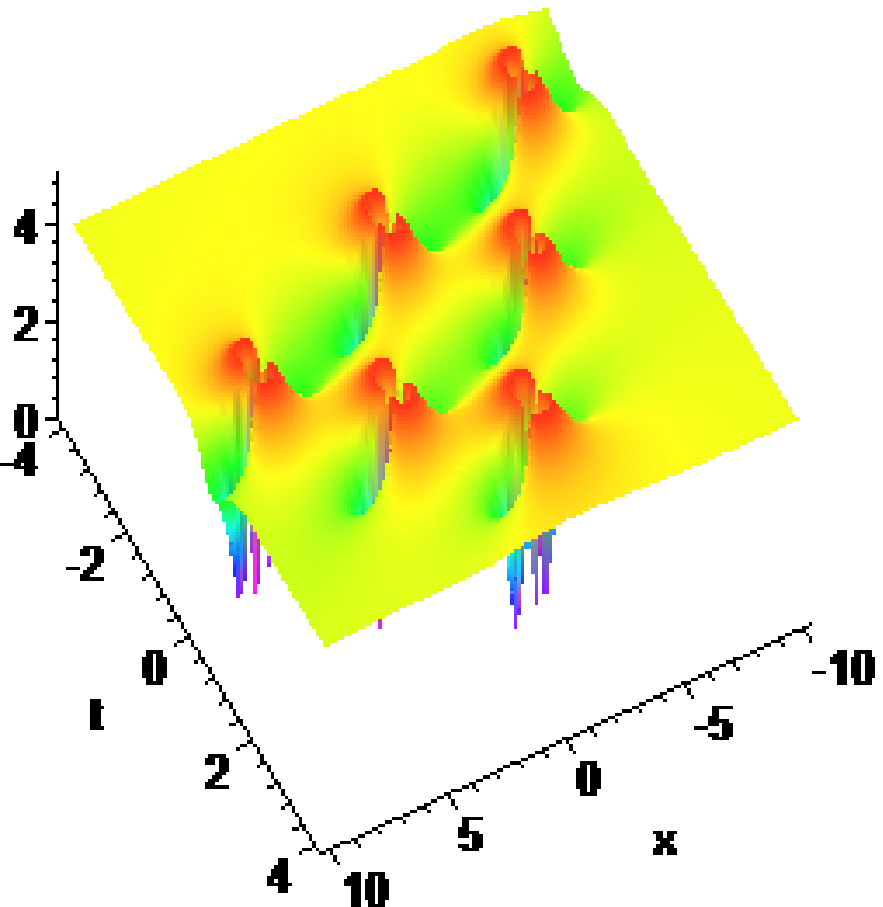,width=4cm}\vspace{0.2cm}\caption{{The third
order dark rogue wave $|\bar {p}^{[3]}|^{2}$ given by
eq.(\ref{rNLS-MBjien2}) for the values used in Figure 19.}}
\end{minipage}
\end{figure}

\begin{figure}[ht]
\setlength{\unitlength}{0.1cm}
\begin{minipage}[t]{8cm}
\epsfig{file=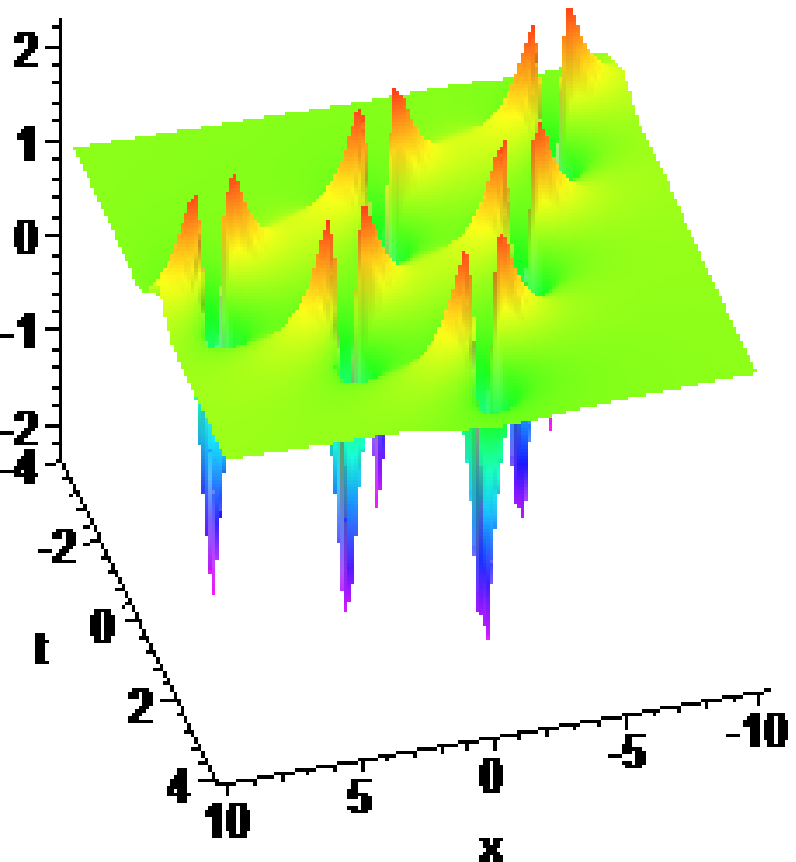,width=4cm}\vspace{0.2cm}\caption{{The third
order dark rogue wave $\bar {\eta}^{[3]} $ given by
eq.(\ref{rNLS-MBjien3}) for the values used in Figure 19. }}
\end{minipage}
\hspace{0.5cm} 
\begin{minipage}[t]{8cm}
\epsfig{file=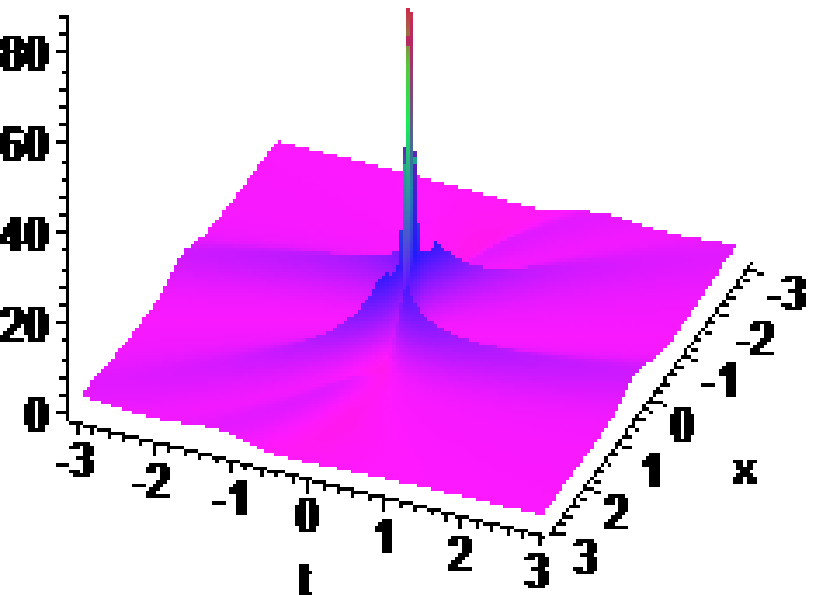,width=4cm}\vspace{0.2cm}\caption{{The third
order rogue wave $|\bar {E}^{[3]}|^{2}$ given by
eq.(\ref{rNLS-MBjien1})with specific parameters $d=\dfrac{4}{3},b=0,
\omega_{0}=\dfrac{1}{2},K_0=1,J_0=0,J_1=0,J_2=0 $. }}
\end{minipage}
\end{figure}

\begin{figure}[ht]
\setlength{\unitlength}{0.1cm}
\begin{minipage}[t]{8cm}
\epsfig{file=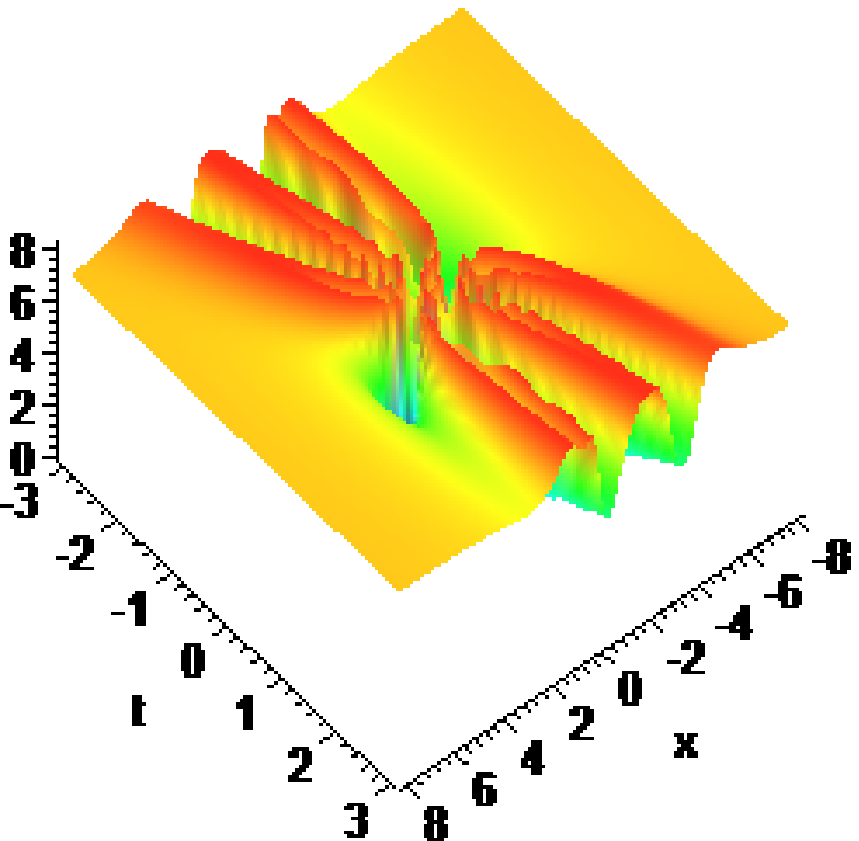,width=4cm}\vspace{0.2cm}\caption{{The third
order dark rogue wave $|\bar {p}^{[3]}|^{2}$ given by
eq.(\ref{rNLS-MBjien2}) for the values used in Figure 22.}}
\end{minipage}
\hspace{0.5cm} 
\begin{minipage}[t]{8cm}
\epsfig{file=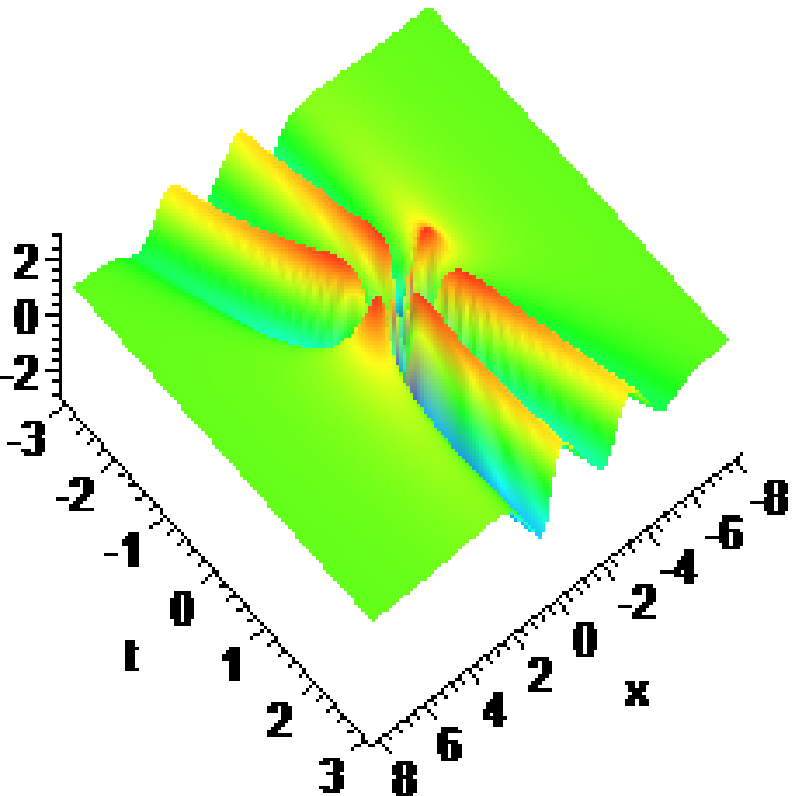,width=4cm}\vspace{0.2cm}\caption{{The third
order dark rogue wave $\bar {\eta}^{[3]} $ given by
eq.(\ref{rNLS-MBjien3}) for the values used in Figure 22. }}
\end{minipage}
\end{figure}


\begin{thebibliography}{99}
\section*{References}
\bibitem{AB1}A.Hasegawa and F.Tappert, Appl.Phys.Lett. 23, 142-144 (1973).
\bibitem{AB2}S.L.McCall and E.L.Hahn, Phys.Rev.Lett. 18, 908-911 (1967).


\bibitem{aim1} A. I. Maimistov A I and E. A. Manykin, Sov. Phys.JETP 58, 685 (1983).
\bibitem{Maimistov}A.I. Maimistov and E. A. Manykin, Sov. Phys. JETP. 85, 1177-1181 (1983).
 \bibitem{aim2} A. I. Maimistov and A. M. Basharov, Nonlinear Optical Waves (Springer-Verlag, Berlin, 1999).


\bibitem{Porsezian1}K. Porsezian and K. Nakkeeran, J. Mod. Opt. 42,1953-1958 (1995).
\bibitem{Kakei} S. Kakei and J. Satsuma, J. Phys. Soc. Japan 63, 885-894 (1994).
\bibitem{Nakazawa1} M. Nakazawa, E. Yamada and H. Kubota, Phys. Rev. Lett. 66,2625-2628(1991).
\bibitem{Nakazawa2} M. Nakazawa, E. Yamada and H. Kubota, Phys.  Rev. A.44, 5973-5987 (1991).

\bibitem{mn1} M.Nakazawa, Y. Kimura, K. Kurokawa  and K.Suzuki , Phys. Rev. A, 45, R23 (1992).

 \bibitem{mn2} M.Nakazawa, K. Suzuki,Y. Kimura and  H. Kubota , Phys. Rev. A 45 R2682  (1992).

\bibitem{Porsezian2}
K. Porsezian and K.Nakkeeran, Phys. Rev. Lett. 74, 2941-2944 (1995).



\bibitem{deformNLSMB}K. Porsezian, P. Seenuvasakumaran,and R. Ganapathy, Phys. Lett. A. 348, 233-243 (2006).
\bibitem{defNLSMB}
A.Mahalingam, K.Porsezian, M.S.Mani Rajan,and A.Uthayakumar, J. Phys. A: Math. Theor.
42, 165101 (2009).
\bibitem{deformHNLSMB}C. G. Latchio Tiofack, Alidou. Mohamadou,Timoleon C Kofane and K. Porsezian, J. Opt. 12, 085202 (2010).
\bibitem{he1}
J.S.He,Y.Cheng and Y.S.Li, Commun. Theor. Phys.38, 493-496 (2002).
\bibitem{GN}G.Neugebauer and R.Meinel, 1984, Phys.Lett.A. 100, 467-470 (1984).
 \bibitem{matveev}V. B. Matveev and M.A. Salle,
 Darboux Transfromations and Solitons (Springer-Verlag, Berlin, 1991).

\bibitem{guo111} Rui.G, Bo.T,  Xing.L,
 Hai-Qiang.Zhang and Wen-Jun.Liu,
 Computational Mathematics and Mathematical Physics, 52, 565-577 (2012).
 \bibitem{Ming} Ming Wang, Wen-Rui Shan, Xing L,
 Bo Qin, and Li-Cai Liu, Z.Naturforsch. 66a, 712-720
 (2011).
\bibitem{Kharif1} C.Kharif and E.Pelinovsky, Eur.J.Mech.B (Fluids). 22, 603-634 (2003).
\bibitem{Kharif2}C.Kharif, E.Pelinovsky and A.Slunyaev, 2009, Rogue Waves in the Ocean (Berlin: Springer).

\bibitem{Akhmediev2}A.Chabchoub, N.P.Hoffmann and N.Akhmediev, Phys. Rev. Lett. 106,204502 (2011).
\bibitem{Didenkulova} I.Didenkulova1 and E.Pelinovsky, Nonlinearity. 24, R1-R18 (2011).
\bibitem{Peregrine}D. H. Peregrine, J. Aust. Math. Soc. Ser. B, Appl. Math. 25, 16-43 (1983).
\bibitem{Dysthe1}Kristian, B.Dysthe and K.Trulsen, Phys Scri. 82, 48-52 (1999).

\bibitem{Zakharov2}V.E.Zakharov and L.A.Ostrovsky, Phys.D. 238, 540 (2009).
\bibitem{Akhmediev3} N.N.Akhmediev and  V.I.Korneev, Theor. Math.
Phys. 69, 1080-1093 (1986).

\bibitem{Shrira} I.Shrira and V.Geogjaev, J.Eng.Math. 67, 11-22 (2010).
\bibitem{yan}Zhenya.Yan, Phys. Lett.A. 374, 672-679 (2009).
\bibitem{Dai}ChaoQing Dai, GuoQuan Zhou and JieFang Zhang, Phys. Rev. E. 85, 016603(2012).

\bibitem{Ying}L.H.Ying,Z.Zhuang,E. J.Heller and L.Kaplan, Nonlinearity. 24, R67-R87 (2011).
 \bibitem{Dubard1}   P. Dubard, P. Gaillard, C. Klein, V.B. Matveev, Eur. Phys. J. Special Topics 185, 247-258 (2010).

\bibitem{Dubard2} P. Dubard, V.B. Matveev, Nat. Hazards. Earth. Syst. Sci. 11, 667-672 (2011).

\bibitem{Gaillard} P. Gaillard, J. Phys. A: Math.Theor. 44, 435204 (2011).

\bibitem{guo} Boling.Guo,Liming.Ling and Q.P.Liu, Phys. Rev. E. 85, 026607(2012).
\bibitem{Yang}Yasuhiro. Ohta, Jianke. Yang, Proc. Proc.R. Soc.A 468, 1716 (2012).

\bibitem{Akhmediev4}N. Akhmediev,J. M. Soto-Crespo and A. Ankiewicz,Phys. Rev. A. 80, 043818 (2009).
\bibitem{JSH1} J. S. He, H. R. Zhang, L. H. Wang, K. Porsezian and
A. S. Fokas, 2012, A generating mechanism for higher order rogue
waves, arXiv:1209.3742v3.
\bibitem{ruderman}V. Fedun, M.S.Ruderman and R. Erd\'elyi,  Phys. Lett.A. 372, 6107-6110 (2008).
\bibitem{derman}
M.S. Ruderman,Eur. Phys. Jour, 185, 57-66 (2010).
\bibitem{xuhe}Shuwei. Xu, Jingsong.He and Lihong.Wang,J. Phys. A: Math. Theor.44, 305203 (2011).

\bibitem{xuhe2}Shuwei. Xu, Jingsong.He and Lihong.Wang, Europhys. Letter.97, 30007 (2012).


\bibitem{Moslem}W.M.Moslem, P.K.Shukla and B.Eliasson, Euro.Phys.Lett. 96, 25002 (2011).
\bibitem{guoliu}
Boling Guo, Liming Ling and Q. P. Liu,   Stud. Appl. Math. (2012) DOI: 10.1111/j.1467-9590.2012.00568.x.

\bibitem{Optical1}
D. R. Solli, C. Ropers, P. Koonath, and B. Jalali,  Nature, 450, 1054-1057 (2007).
\bibitem{Optical2}B. Kibler, J. Fatome, C. Finot, G. Millot,F. Dias, G. Genty,N. Akhmediev and J. M. Dudley, Nature.Physics. 6, 790-795 (2010).
\bibitem{Optical3}F. T. Arecchi, U. Bortolozzo,A. Montina and S. Residori, Phys. Rev. Lett. 106, 153901 (2011).
\bibitem{Akhmediev5}A.Ankiewicz, J. M. Soto-Crespo and N. Akhmediev, Phys. Rev. E. 81, 046602 (2010).
\bibitem{Guangye} Guangye Yang, Lu Li and Suotang Jia, Phys. Rev. E. 85, 046608(2012).
\bibitem{taohe1} Yongsheng Tao and  Jingsong He,  Phys. Rev. E 85, 026601 (2012).
\bibitem{Akhmediev6}A.Ankiewicz, N.Akhmediev and J. M. Soto-Crespo, Phys. Rev. E. 82, 026602 (2010).

\bibitem{GuoBL2} Boling.Guo and Liming.Ling, Chin. Phys. Lett.  28, 110202 (2011).
\bibitem{Baronio}  Fabio Baronio, Antonio Degasperis, Matteo Conforti and Stefan Wabnitz, 2012, Phys. Rev. Lett. 109, 044102(2012).


\bibitem{WeiGuo}
BaoGuo.Zhai, WeiGuo. Zhang, XiaoLiWang and HaiQiang Zhang,
Nonlinear Anal. RWA 14, 14-27 (2012).



\bibitem{Zhenyun}
Zhenyun Qin and Gui Mu, Phys.  Rev. E. 86, 036601(2012).
\bibitem{Rider}
Rider. Jaimes-Re\'ategui, Ricardo. Sevilla-Escoboza and G.Huerta-Cuellar, Phys.Rev. Lett. 107, 274101 (2011).

\bibitem{Zakharov} V.E.Zakharov and A.B.Shabat, Soviet.Phys.JETP.37, 823-828(1973).

\bibitem{NewTypes}
Jingsong. He, Shuwei. Xu and K. Porsezian, J. Phys. Soc. Jpn.  81, 033002
(2012).
\bibitem{he2} J.S.He, L.Zhang, Y.Cheng and Y.S.Li, Science in China Series A:
Mathematics.12, 1867-1878 (2006).
\bibitem{triplets}A. Ankiewicz, J.D. Kedziora, N. Akhmediev, Phys. Lett.A. 375, 2782-2785 (2011).
\bibitem{Circular}
David.J. Kedziora, Adrian. Ankiewicz, and Nail. Akhmediev, Phys. Rev. E. 84, 056611 (2011).
\bibitem{scg} J.M.Dudley and J.R.Taylor (Eds.), Supercontinuum Generation in Optical fibres(Cambridge,Cambridge University
Press, 2010).
\bibitem{kpMI} B.Kalithasan, K.Porsezian,  P.T.Dinda  and B.A.Malomed,  J. Optics A:  Pure and Applied Optics 11, 045205 (2009).
\bibitem{kp1} K.Porsezian,A.Hasegawa, V.N.Serkin, T.L.
Belyaeva and R.Ganapathy, Phys. Lett. A., 361, 504-508 (2007).

\bibitem{serkin1} V. N. Serkin and A. Hasegawa, Phys. Rev. Lett. 85,
 4502 (2000).
\bibitem{serkin2} V. N. Serkin, A. Hasegawa,
and T. L. Belyaeva, Phys. Rev. Lett. 98, 074102 (2007).
\bibitem{serkin3} V.N. Serkin, A. Hasegawa, T.L. Belyaeva,
 Journal of Modern Optics 57 , 1456 (2010).
\bibitem{JSH2} Chuanzhong Li, Jingsong He and K. Porsezian,
arXiv:1205.1191v1 (2012).
\bibitem{liheuu} Chuanzhong Li and Jingsong He,
arXiv:1210.2501 (2012).
\bibitem{shanxue} Yu-Shan Xue, Bo Tian, Wen-Bao Ai, Min Li and Pan Wang
Optics \& Laser Technology, \textbf{48}, 153-159(2013).

\end{thebibliography}
\end{document}